\documentclass[12pt]{article}
\usepackage[utf8]{inputenc}
\usepackage{geometry}
\usepackage{float}
\usepackage{amsmath}
\usepackage{amssymb}
\usepackage{graphicx}
\usepackage{setspace}
\usepackage[authoryear]{natbib}
\usepackage{amsthm}
\usepackage[titletoc,title]{appendix}
\usepackage{authblk}
\usepackage{color}
\usepackage{rotating}

\makeatletter

\providecommand{\tabularnewline}{\\}

\usepackage{latexsym}
\usepackage{amsthm}\usepackage{amsfonts}\usepackage{graphicx}\usepackage{epsfig}
\usepackage{bm}
\newcommand*{\patchAmsMathEnvironmentForLineno}[1]{%
      \expandafter\let\csname old#1\expandafter\endcsname\csname #1\endcsname
      \expandafter\let\csname oldend#1\expandafter\endcsname\csname end#1\endcsname
      \renewenvironment{#1}%
         {\linenomath\csname old#1\endcsname}%
         {\csname oldend#1\endcsname\endlinenomath}}%
    \newcommand*{\patchBothAmsMathEnvironmentsForLineno}[1]{%
      \patchAmsMathEnvironmentForLineno{#1}%
      \patchAmsMathEnvironmentForLineno{#1*}}%
    \AtBeginDocument{%
    \patchBothAmsMathEnvironmentsForLineno{equation}%
    \patchBothAmsMathEnvironmentsForLineno{align}%
    \patchBothAmsMathEnvironmentsForLineno{flalign}%
    \patchBothAmsMathEnvironmentsForLineno{alignat}%
    \patchBothAmsMathEnvironmentsForLineno{gather}%
    \patchBothAmsMathEnvironmentsForLineno{multline}%
    }
\usepackage[mathlines]{lineno}

\include{def}
\def\dispmuskip{\thinmuskip= 3mu plus 0mu minus 2mu \medmuskip=  4mu plus 2mu minus 2mu \thickmuskip=5mu plus 5mu minus 2mu}
\def\textmuskip{\thinmuskip= 0mu                    \medmuskip=  1mu plus 1mu minus 1mu \thickmuskip=2mu plus 3mu minus 1mu}
\def\beq{\dispmuskip\begin{equation}}    \def\eeq{\end{equation}\textmuskip}
\def\beqn{\dispmuskip\begin{displaymath}}\def\eeqn{\end{displaymath}\textmuskip}
\def\bea{\dispmuskip\begin{eqnarray}}    \def\eea{\end{eqnarray}\textmuskip}
\def\bean{\dispmuskip\begin{eqnarray*}}  \def\eean{\end{eqnarray*}\textmuskip}

\newtheorem{lemma}{Lemma}

\newtheorem{example}{Example}

\newcommand{\bx}{\bm{x}}
\newcommand{\bu}{\bm{u}}

\def\E{{\mathbb E}}                         

\def\D{{\cal D}}

\def\C{{\cal C}}
\def\D{{\cal D}}

\def\J{{\cal J}}
\def\M{{\cal M}}

\def\bsX{\ensuremath{\boldsymbol{X}}}
\def\bsU{\ensuremath{\boldsymbol{U}}}
\def\bsx{\ensuremath{\boldsymbol{x}}}
\def\bsu{\ensuremath{\boldsymbol{u}}}

\def\mrd{{\ensuremath{\mathrm{d}}}}

\makeatother

\catcode`\<=\active \def<{
\fontencoding{T1}\selectfont\symbol{60}\fontencoding{\encodingdefault}}
\catcode`\>=\active \def>{
\fontencoding{T1}\selectfont\symbol{62}\fontencoding{\encodingdefault}}
\newcommand{\mathd}{\mathrm{d}}

\newcommand{\tmmathbf}[1]{\ensuremath{\boldsymbol{#1}}}

\newcommand{\mathbbm}[1]{\text{\usefont{U}{bbm}{m}{n}#1}} 

\author[1]{David Gunawan}
\author[2]{Mohamad A. Khaled}
\author[1]{Robert Kohn}
\affil[1]{University of New South Wales and ACEMS}
\affil[2]{University of Queensland}

\renewcommand\Authands{ and }

\begin{document}

\title{Mixed Marginal Copula Modeling}
\maketitle
\begin{abstract}
This article extends the literature on copulas with discrete or continuous marginals to the case where
some of the marginals are a mixture of discrete and continuous components. We do so  by carefully defining
the likelihood as the density of the observations with respect to a mixed measure. The treatment is quite general, although we focus focus on
mixtures of Gaussian and Archimedean copulas. The inference is Bayesian with the estimation carried out by Markov chain Monte Carlo.
We illustrate the methodology and algorithms by applying them to
estimate a multivariate income dynamics model.

\end{abstract}

\section{Introduction}\label{S: introduction}
Copulas are a versatile and useful tool for
modeling multivariate distributions.
See, for example, \cite{fan2014copulas}, \cite{patton2009copula},
\cite{durante2015principles} and \cite{trivedi2007copula}.
Modeling non-continuous marginal
random variables is  a challenging task due to computational
problems, interpretation difficulties and various other pitfalls and
paradoxes; see \cite{Smith2012}, for example. \textcolor{blue}{The main source of the
computational issues arises from the difficulty of directly evaluating the likelihood. For example, when modeling a vector of $m$ discrete random variables, evaluating the likelihood at one point requires computing $2^m$ terms.} The literature on modeling non-continuous random marginal problems
has mostly focused on cases where all the marginals are discrete, and less
extensively, on cases where some marginals are discrete and some are continuous.
See, for example,
\cite{genest2007primer}, \cite{Smith2012}, \cite{de2013analysis},
and \cite{panagiotelis2012pair}.
Furthermore, a lot of the literature has focused on approaches restricted to certain classes to copulas. For example, this is the case for Gaussian copulas (See for instance \cite{shen2006copula}, \cite{hoff2007extending}, \cite{song2009joint}, \cite{de2011copula}, \cite{he2012gaussian} and
\cite{jiryaie2016gaussian}) or pair-copula constructions (see \cite{stober2015comorbidity}). Relatively little attention has been paid to the case where some variables are a mixture of discrete and continuous components.
In contrast, our approach, presents methodology for an arbitrary copula and can be applied quite generally as long as it is possible to compute certain marginal and conditional copulas either in closed-form or numerically.

Our article extends the Bayesian methodology used
for estimating continuous marginals to the case where each marginal can be
a mixture of an absolutely continuous random variable and a discrete random variable.
In particular, we are interested in applying the new methodology to copulas that are mixtures of Gaussian and Archimedean copulas.
To illustrate the methodology and sampling algorithm we apply them to estimate a multivariate income dynamics model.
In this application, we use the copula framework to
model the dependence structure of random variables that are mixtures of discrete and continuous components, and apply the model
to empirical economic data.
We note that there are many other
real world economic applications that involve such mixtures of random variables as marginals, and these are briefly discussed in Section~\ref{S: conclusion}.

Our proposed methodology extends  that introduced in \cite{Pitt2006} and \cite{Smith2012}. \cite{Smith2012} allow  the joint modeling of
distributions of random variables such that each component can be either discrete or continuous. However, neither paper covers the case where some random variables can be
 a mixture of an absolutely continuous random variable and a discrete random variable. In a financial econometrics application, \cite{brechmann2014}
consider the case where the marginal distributions are mixtures of continuous and \textcolor{blue}{points of probability mass at zero}.
In contrast, our paper derives the likelihood equations in a much more general setting that allows for the
margins to be arbitrarily classified into three groups: absolutely continuous, discrete and mixtures of absolutely continuous and discrete random variables. Furthermore, there is
no restriction on the number or location of the point masses present in each margin. This can occur in many economic data, for instance in cases where earnings are top-coded and have individuals with zero earnings. Equally, our setting covers the case of dependent interval-censored data.

The paper is organized as follows. Section~\ref{S: likelihood definition} outlines the copula model and defines the likelihood as a density with respect to a mixed measure.
Section~\ref{S: estimation and algorithms} presents the simulation algorithms used for inference.
Section~\ref{S: application to individual income dynamics} applies the methods and algorithms
to model multivariate income dynamics. This section describes the data and presents the estimation results.
Section~\ref{S: conclusion} concludes. \textcolor{blue}{The paper has two appendices}.
Appendix~\ref{app_diference_operator} defines the \textcolor{blue}{difference notation which is a handy tool useful when writing
formulas for the likelihood of our model in closed-form.} Appendix \ref{app_integration} presents and proves the results required to define the likelihood
as a density with respect to a mixed measure. The paper also has an online supplement whose sections are denoted as Sections~S1, etc.
Section~\ref{app_sampling_schemes} describes the Gaussian and Archimedean copulas used in the article, as well as
 the Markov chain Monte Carlo (MCMC) sampling scheme.  \textcolor{blue}{Section~\ref{S: trivariate example}
 introduces a new three dimensional example to further illustrate the methods in the paper.
Section~\ref{proof: proof of lemma 3} gives a proof of Lemma~\ref{L: elementary} which is discussed in
 Appendix~\ref{app_integration}}.  Section~\ref{app: extra empirical results} presents some additional empirical results.

\section{Defining the Likelihood of a general copula\label{S: likelihood definition}}

This section discusses the proposed model and shows how to write the likelihood of an i.i.d. sample from it.
Each random vector is modeled using a marginal
distribution-copula decomposition and each marginal is allowed to be a mixture
of an absolutely continuous component  and \textcolor{blue}{ a discrete component}. The MCMC sampling scheme
in the next section is based on this definition of the likelihood.

Let $\ensuremath{\boldsymbol{X}}= (X_1, \ldots, X_m)$ be \textcolor{blue}{an $\mathbbm{R}^m$-valued random vector.
If, for example, $X_j$ is categorical, then
its support would be a finite subset of $\mathbbm{R}$ and thus without loss of generality, we can work with $\mathbbm{R}^m$.} Let
$\mathcal{M}= \{ 1, \ldots, m \}$ be the index set, and $2^{\mathcal{M}}$ its
power-set (or the set of all of its subsets). Let the random variable $X_j$
have cumulative distribution function $F_j$ for $j = 1, \ldots, m$. By the
Lebesgue decomposition theorem \textcolor{blue}{\citep[][Chapter 7, Theorem 1.1]{shorack2000probability},
and assuming there are  no continuous singularities \citep[see][for a detailed discussion]{durante2015principles}},
the distribution of each $X_j$ can be written as a mixture
of an absolutely continuous random variable and a discrete random variable.
This means that $F_j$ is allowed to have jumps at a countable number of
points. In order to exploit this result, we would like to be able to
decide at each point of \textcolor{blue}{$\mathbbm{R}^m$}, which indices have jumps in
their corresponding CDFs.

We need a mapping $\mathcal{C}: \mathbbm{R}^m \rightarrow 2^{\mathcal{M}}$
that, for each $\ensuremath{\boldsymbol{x}} \in \mathbbm{R}^m$, picks out the
subset of the indices of $\ensuremath{\boldsymbol{x}}$ where $F_j$ is
continuous at $x_j$ for each $j \in \mathcal{C}
(\ensuremath{\boldsymbol{x}})$.
\begin{align*}
     \mathcal{C} : \mathbbm{R}^m  & \longrightarrow  2^{\mathcal{M}} \quad \text{with} \quad
      \ensuremath{\boldsymbol{x}} \longrightarrow  \mathcal{C}
     (\ensuremath{\boldsymbol{x}}).
\end{align*}
Similarly, we define the set $\mathcal{D} (\ensuremath{\boldsymbol{x}}) = \mathcal{M} -
 \mathcal{C} (\ensuremath{\boldsymbol{x}})$ (the complement of
$\mathcal{C} (\ensuremath{\boldsymbol{x}})$ in $\mathcal{M}$, that is the
set of indices $j$ for which $F_j$ presents jumps at $x_j$). This means
that for all $ \ensuremath{\boldsymbol{x}} \in \mathbbm{R}^m$, $\{
\mathcal{C} (\ensuremath{\boldsymbol{x}}), \mathcal{D}
(\ensuremath{\boldsymbol{x}}) \}$ partitions the index set so that
$\mathcal{C} (\ensuremath{\boldsymbol{x}}) \cap \mathcal{D}
(\ensuremath{\boldsymbol{x}}) = \varnothing$ and $\mathcal{C}
(\ensuremath{\boldsymbol{x}}) \cup \mathcal{D} (\ensuremath{\boldsymbol{x}})
=\mathcal{M}$.

As a first example, consider $\ensuremath{\boldsymbol{X}}= (X_1, X_2)$, where
$X_1 \sim \mathcal{N} (0, 1)$ and $X_2$ is a mixture of an exponential distribution with
parameter $\lambda$ and a point mass at $0$ with probability $p$, i.e., $X_2
\sim p \delta_0 + (1 - p) \mathcal{E} (\lambda)$). Then, $\mathcal{C} (x_1, 0) = \{ 1 \}$ for all
$ x_1 \in \mathbbm{R}$
and $\mathcal{C} (x_1, x_2) = \{ 1, 2 \}$ for all $x_1 \in \mathbbm{R}, x_2 >
0$. Similarly $\mathcal{D} (x_1, 0) = \{ 2 \}$ for all $x_1 \in \mathbbm{R}$
and $\mathcal{D} (x_1, x_2) = \varnothing$.

As a second example, let $\ensuremath{\boldsymbol{X}}= (X_1, X_2)$,  where $X_1$
is Bernoulli and $X_2 \sim \mathcal{N} (0, 1)$. Then $\mathcal{C}
(\ensuremath{\boldsymbol{x}}) = \{ 2 \}$ for all $ \ensuremath{\boldsymbol{x}}
\in \{ 0, 1 \} \times \mathbbm{R}$. Similarly $\mathcal{D}
(\ensuremath{\boldsymbol{x}}) = \{ 1 \}$ for all $
\ensuremath{\boldsymbol{x}}$.

Let $\ensuremath{\boldsymbol{U}}= (U_1, \ldots, U_m)$ be a vector of
uniform random variables whose distribution is given by some copula $C$.
We assume that $F_j^{- 1}$ is the quantile function corresponding to $F_j$
(since $F_j$ is not invertible when \textcolor{blue}{ $X_j$ is not absolutely continuous}, this
corresponds to picking one possible generalized inverse function).

The variables $\ensuremath{\boldsymbol{U}}$ are selected to satisfy
the following criteria. If, at coordinate $x_j$, $j \in \mathcal{C}
(\ensuremath{\boldsymbol{x}})$, then $u_j = F_j (x_j)$, resulting in
a deterministic one-to-one relationship when conditioning on either $U_j$ or
$X_j$. Otherwise,  $j \in \mathcal{D} (\ensuremath{\boldsymbol{x}})$, and
we require $x_j = F_j^{- 1} (u_j)$, resulting in an infinity of
$U_j$ corresponding to one $X_j$ and spanning the interval $(F_j (X_j^-), F_j
(X_j))$. This interval corresponds to gaps in the range of $F_j$. If
$\mathcal{C} (\ensuremath{\boldsymbol{x}}) =\mathcal{M}$ for every
$\ensuremath{\boldsymbol{x}}$, then $C$ will be the copula of
$\ensuremath{\boldsymbol{X}}$. Otherwise, the copula structure will still
create dependence between \textcolor{blue}{the non-continuous marginal variables but will not be unique in
general}.
Mathematically, the above description leads to the joint density
\begin{equation}
  \label{JointDensity} f (\ensuremath{\boldsymbol{x}},
  \ensuremath{\boldsymbol{u}}) := c (\ensuremath{\boldsymbol{u}}) \prod_{j \in
  \mathcal{C} (\ensuremath{\boldsymbol{x}})} \mathcal{I} (u_j = F_j (x_j))
  \prod_{j' \in \mathcal{D} (\ensuremath{\boldsymbol{x}})} \mathcal{I} (F_{j'}
  (x_{j'}^-) \leqslant u_{j'} < F_{j'} (x_{j'})),
\end{equation}
where $c$
is the copula density corresponding to $C$ and $\mathcal{I}$ is an indicator
variable. See Lemma~\ref{L: mixed density in x and u}, part~(i), of Appendix~~\ref{app_integration} for a derivation of
\eqref{JointDensity} and the corresponding measure.
Notice that in ~\eqref{JointDensity}, products over the indices $j$ and $ j'$ correspond to different
partitions for each $\ensuremath{\boldsymbol{x}}$.

With a small abuse of notation, we call $\ensuremath{\boldsymbol{U}}$ the vector of latent variables, even though $U_j$ is a deterministic function of
$X_j$ if $F_j$ is invertible.

To derive the likelihood function, that is the marginal density of
$\ensuremath{\boldsymbol{X}}$, from the joint density $f (\ensuremath{\boldsymbol{x}},
\ensuremath{\boldsymbol{u}})$, we introduce some notation. Let
$\ensuremath{\boldsymbol{a}}, \ensuremath{\boldsymbol{b}}$ be two vectors in
$\mathbbm{R}^k$ such that $\ensuremath{\boldsymbol{a}} \leqslant
\ensuremath{\boldsymbol{b}}$ componentwise and let $g$ be an arbitrary
function from $\mathbbm{R}^k$ into $\mathbbm{R}$. \textcolor{blue}{We denote by
$\bigtriangleup_{\ensuremath{\boldsymbol{a}}}^{\ensuremath{\boldsymbol{b}}} g
(.)$ the sum} of $2^k$ terms that are obtained by repeatedly subtracting $g
(., a_j, .)$ from $g (., b_j, .)$ for each $j = 1, \ldots, k$. Appendix
\ref{app_diference_operator} contains more details on using this notation.

For each $\ensuremath{\boldsymbol{x}} \in \mathbbm{R}^m$, denote by
$\ensuremath{\boldsymbol{b}}= (F_1 (x_1), \ldots, F_m (x_m))$ the vector of upper bounds and similarly denote by
$\ensuremath{\boldsymbol{a}}= (F_1 (x_1^-), \ldots, F_m (x_m^-))$ the vector of lower bounds. For each $j
\in \mathcal{C} (\ensuremath{\boldsymbol{x}})$, $\ensuremath{\boldsymbol{b}}
(j) =\ensuremath{\boldsymbol{a}} (j)$, otherwise we have the strict inequality
$\ensuremath{\boldsymbol{b}} (j) >\ensuremath{\boldsymbol{a}} (j)$. Denote the
partitions of $\ensuremath{\boldsymbol{a}}$ and $\ensuremath{\boldsymbol{b}}$
by $\ensuremath{\boldsymbol{a}}_{\mathcal{C} (\ensuremath{\boldsymbol{x}})}$,
$\ensuremath{\boldsymbol{a}}_{\mathcal{D} (\ensuremath{\boldsymbol{x}})}$,
$\ensuremath{\boldsymbol{b}}_{\mathcal{C} (\ensuremath{\boldsymbol{x}})}$ and
$\ensuremath{\boldsymbol{b}}_{\mathcal{D} (\ensuremath{\boldsymbol{x}})}$. For
some sets $A, B \subset \mathcal{M}$, denote by $c_A$ and $c_{A|B}$, \ the
marginal copula density over the indices of $A$, the conditional copula
density where the variables in $A$ are conditioned on the variables
with index set $B$. It is possible to do the same for \textcolor{blue}{ $C_A$ and $C_{A|B}$, the copula
distribution functions.}

If $(\ensuremath{\boldsymbol{X}}, \ensuremath{\boldsymbol{U}})$ has the joint
density given by \eqref{JointDensity}, then the marginal density of
$\ensuremath{\boldsymbol{X}}$ is
\begin{equation}
  \label{likelihood} f (\ensuremath{\boldsymbol{x}}) = c_{\mathcal{C}
  (\ensuremath{\boldsymbol{x}})} (\ensuremath{\boldsymbol{b}}_{\mathcal{C}
  (\ensuremath{\boldsymbol{x}})}) \prod_{j \in \mathcal{C}
  (\ensuremath{\boldsymbol{x}})} f _j(x_j)
  \bigtriangleup_{\ensuremath{\boldsymbol{a}}_{\mathcal{D}
  (\ensuremath{\boldsymbol{x}})}}^{\ensuremath{\boldsymbol{b}}_{\mathcal{D}
  (\ensuremath{\boldsymbol{x}})}} C_{\mathcal{D} (\ensuremath{\boldsymbol{x}})
  |\mathcal{C} (\ensuremath{\boldsymbol{x}})}
  (\cdot|\ensuremath{\boldsymbol{b}}_{\mathcal{C} (\ensuremath{\boldsymbol{x}})}),
\end{equation}
which corresponds to writing the formula for the density of
$\ensuremath{\boldsymbol{X}}$ as the product of the (marginal) density of
continuous components at $\ensuremath{\boldsymbol{x}}$
\[ f (\ensuremath{\boldsymbol{x}}_{\mathcal{C} (\ensuremath{\boldsymbol{x}})})
   = c_{\mathcal{C} (\ensuremath{\boldsymbol{x}})}
   (\ensuremath{\boldsymbol{b}}_{\mathcal{C} (\ensuremath{\boldsymbol{x}})})
   \prod_{j \in \mathcal{C} (\ensuremath{\boldsymbol{x}})} f _j(x_j),  \]
and the (conditional) density of the non-continuous components conditional on
the continuous ones
\[ f (\ensuremath{\boldsymbol{x}}_{\mathcal{D} (\ensuremath{\boldsymbol{x}})}
   |\ensuremath{\boldsymbol{x}}_{\mathcal{C} (\ensuremath{\boldsymbol{x}})}) =
   \bigtriangleup_{\ensuremath{\boldsymbol{a}}_{\mathcal{D}
   (\ensuremath{\boldsymbol{x}})}}^{\ensuremath{\boldsymbol{b}}_{\mathcal{D}
   (\ensuremath{\boldsymbol{x}})}} C_{\mathcal{D}
   (\ensuremath{\boldsymbol{x}}) |\mathcal{C} (\ensuremath{\boldsymbol{x}})}
   (\cdot|\ensuremath{\boldsymbol{b}}_{\mathcal{C} (\ensuremath{\boldsymbol{x}})}).
\]
See Lemma~\ref{L: mixed density in x and u}, part~(ii), of Appendix~~\ref{app_integration} for a derivation of
\eqref{likelihood} and the corresponding measure.

\textcolor{blue}{We now give a bivariate example to illustrate how the formulas can be used. This example is continued in later
sections. See also Section~\ref{S: trivariate example} for a trivariate illustrative example. }

\begin{example}[running illustrative example]\label{ex: example 1}

Let $X_1$ have a density that is a mixture of point of probability mass at
zero and a normal distribution $f_1 (x_1) \sim \pi \delta_{x_1} (0) + (1 -
\pi) \phi (x_1)$ where $\phi (.)$ is the density of a standard normal.
This implies that the cumulative distribution function of $X_1$ is
\[ F_1 (x_1) = (1 - \pi) \Phi (x_1) + \pi \mathcal{I} (x_1 \geqslant 0), \]
and thus there a discontinuity in $F_1$ at the point 0.
Let $X_2$ be a binary random variable with $\Pr \{ X_2 = 0 \} = \gamma$.

Let $C (\cdot)$ and $c (\cdot)$ be respectively the Clayton copula and
Clayton copula density with parameter $\theta = 1$, so that
\[ C (u_1, u_2) = \left( \frac{1}{u_1} + \frac{1}{u_2} - 1 \right)^{- 1} , \quad
 c (u_1, u_2) = \frac{2}{u_1^2 u_2^2} \left( \frac{1}{u_1} + \frac{1}{u_2} -
   1 \right)^{- 3} \]
and the conditional copula is given by
\[ C_{2|1} (u_2 |u_1) = \frac{1}{u_1^2} \left( \frac{1}{u_1} + \frac{1}{u_2} -
   1 \right)^{- 2}, \]
which has the conditional quantile function $C^{- 1} (\tau |u_1) =
\frac{\sqrt{\tau} u_1}{1 + \sqrt{\tau} (u_1 - 1)}$ and the conditional density
$c_{2|1} (u_2 |u_1) = c (u_1, u_2)$ (because the marginal of $u_1$ is
uniform).

The following details are necessary construct the example.

$\mathcal{C} (\tmmathbf{x}) = \{ 2 \}$ for $x_1 \neq 0$, for all $ x_2$ and
$\mathcal{C} (\tmmathbf{x}) = \{ 1, 2 \}$ for $x_1 = 0$, for all $ x_2$

{\underline{Joint of $\tmmathbf{x}$ and $\tmmathbf{u}$}} (  Eq.~\eqref{JointDensity} )

There are two cases. Case 1:  $x_1 \neq 0$

$f (x_1, x_2, u_1, u_2) = c (u_1, u_2) \mathcal{I} (u_1 = F_1 (x_1))
\mathcal{I} (F_2 (x_2 -) \leqslant u_2 < F_2 (x_2))$

Case 2: $x_1 = 0$

$f (x_1, x_2, u_1, u_2) = c (u_1, u_2) \mathcal{I} (F_1 (0 -) \leqslant u_1 <
F_1 (0)) \mathcal{I} (F_2 (x_2 -) \leqslant u_2 < F_2 (x_2))$

{\underline{Likelihood at one point}} (Eq.~\ref{likelihood} )

If $x_1 \neq 0$, then
\begin{eqnarray*}
  f (x_1, x_2) & = & f (x_1) \bigtriangleup_{F_2 (x_2 -)}^{F_2 (x_2)} C_{2|1}
  (\cdot |F (x_1))\\
  & = & f_1 (x_1) \{ C_{2|1} (F_2 (x_2) |F_1 (x_1)) - C_{2|1} (F_2 (x_2 -)
  |F_1 (x_1)) \}
\end{eqnarray*}
because $c (u_1) = 1$ as one-dimensional margins of a copula
are all uniform.
If $x_1 = 0$, then
\begin{eqnarray*}
  f (0, x_2) & = & \bigtriangleup_{F_1 (0 -)}^{F_1 (0)} \bigtriangleup_{F_2
  (x_2 -)}^{F_2 (x_2)} C (\cdot)\\
  & = & \bigtriangleup_{F_1 (0 -)}^{F_1 (0)} \{ C (\cdot, F_2 (x_2)) - C
  (\cdot, F_2 (x_2 -)) \}\\
  & = & C (F_1 (0), F_2 (x_2)) - C (F_1 (0), F_2 (x_2 -)) - C (F_1 (0 -), F_2
  (x_2)) + C (F_1 (0 -), F_2 (x_2 -)).
\end{eqnarray*}

\end{example}

The difficult part of implementing a simulation algorithm based on  equations \eqref{JointDensity} and
\eqref{likelihood} is that the size of the vectors
$\ensuremath{\boldsymbol{x}}_{\mathcal{C} (\ensuremath{\boldsymbol{x}})}$ and
$\ensuremath{\boldsymbol{x}}_{\mathcal{D} (\ensuremath{\boldsymbol{x}})}$
changes with $\ensuremath{\boldsymbol{x}}$. A secondary  difficulty is that the
second term is a sum of $2^{| \mathcal{D} (\ensuremath{\boldsymbol{x}}) |}$ terms
for each $\ensuremath{\boldsymbol{x}}$, where $| \mathcal{D}
(\ensuremath{\boldsymbol{x}}) |$ is the cardinality of the set $\mathcal{D}
(\ensuremath{\boldsymbol{x}})$.

\section{Estimation and Algorithms\label{S: estimation and algorithms}}

\subsection{Conditional distribution of the latent variables\label{SS: conditional distn of latent variables}}

In any simulation scheme (such as MCMC or simulated EM) where the latent
variables $\ensuremath{\boldsymbol{U}}$ are used to carry out inference, it
is  necessary to know the distribution of
$\ensuremath{\boldsymbol{U}}|\ensuremath{\boldsymbol{X}}$. This distribution
is singular due to the deterministic relationship over
$\mathcal{C} (\ensuremath{\boldsymbol{x}})$ for each
$\ensuremath{\boldsymbol{x}} \in \mathbbm{R}^m$. For this reason, it is useful
to work only with $\ensuremath{\boldsymbol{U}}_{\mathcal{D}
(\ensuremath{\boldsymbol{x}})} |\ensuremath{\boldsymbol{X}}$. A second  issue is
the  need to work with different sizes of vectors
$\ensuremath{\boldsymbol{U}}_{\mathcal{D} (\ensuremath{\boldsymbol{x}})}$ for
each $\ensuremath{\boldsymbol{x}}$ in our sample (say
$\ensuremath{\boldsymbol{x}}_1, \ldots, \ensuremath{\boldsymbol{x}}_n$), so we
will be working with $n$ distributions over different spaces. Recursively
using Bayes formula and similar integration arguments to the ones described
during the derivation of the $\ensuremath{\boldsymbol{X}}$ density, we obtain
the density for $\ensuremath{\boldsymbol{U}}_{\mathcal{D}
(\ensuremath{\boldsymbol{x}})} |\ensuremath{\boldsymbol{X}}$ as
\begin{equation}
  \label{LatentConditionalDistribution} f
  (\ensuremath{\boldsymbol{u}}_{\mathcal{D} (\ensuremath{\boldsymbol{x}})}
  |\ensuremath{\boldsymbol{x}}) = \frac{c_{\mathcal{D}
  (\ensuremath{\boldsymbol{x}}) |\mathcal{C} (\ensuremath{\boldsymbol{x}})}
  (\ensuremath{\boldsymbol{u}}_{\mathcal{D} (\ensuremath{\boldsymbol{x}})}
  |\ensuremath{\boldsymbol{b}}_{\mathcal{C} (\ensuremath{\boldsymbol{x}})})
  \prod_{j \in \mathcal{D} (\ensuremath{\boldsymbol{x}})} \mathcal{I} (a_j
  \leqslant u_j <
  b_j)}{\bigtriangleup_{\ensuremath{\boldsymbol{a}}_{\mathcal{D}
  (\ensuremath{\boldsymbol{x}})}}^{\ensuremath{\boldsymbol{b}}_{\mathcal{D}
  (\ensuremath{\boldsymbol{x}})}} C_{\mathcal{D} (\ensuremath{\boldsymbol{x}})
  |\mathcal{C} (\ensuremath{\boldsymbol{x}})}
  (\cdot|\ensuremath{\boldsymbol{b}}_{\mathcal{C} (\ensuremath{\boldsymbol{x}})})},
\end{equation}
where the denominator is a constant of integration. As seen from the
above conditional density, one of the complexities arising is that the
distribution $\ensuremath{\boldsymbol{U}}_{\mathcal{D}
(\ensuremath{\boldsymbol{x}})}
|\ensuremath{\boldsymbol{X}}=\ensuremath{\boldsymbol{x}}$ depends on the whole
vector $\ensuremath{\boldsymbol{x}}$ and not just on
$\ensuremath{\boldsymbol{x}}_{\mathcal{D} (\ensuremath{\boldsymbol{x}})}$.
See Lemma~\ref{L: mixed density in x and u}, part~(iii), of Appendix~~\ref{app_integration} for a derivation of
\eqref{LatentConditionalDistribution} and the corresponding measure.

We can now proceed in two ways. We can either draw each $U_j$ in $\ensuremath{\boldsymbol{U}}_{\mathcal{D}(\ensuremath{\boldsymbol{x}})}$ separately conditionally on everything else. This is reminiscent of a single move Gibbs sampler. Alternatively, it turns out that in spite of the difficulties, the above distribution can also
be sampled recursively without having to compute any of the above normalizing
constants. By writing $\mathcal{D} (\ensuremath{\boldsymbol{x}})$  as
$\{ j_1, \ldots, j_{| \mathcal{D} (\ensuremath{\boldsymbol{x}}) |} \}$,
we can use the following scheme
\begin{itemize}
  \item $U_{j_1} |\ensuremath{\boldsymbol{X}}$

  \item $U_{j_2} |U_{j_1}, \ensuremath{\boldsymbol{X}}$

  \item $\vdots$

  \item $U_{j_{| \mathcal{D} (\ensuremath{\boldsymbol{x}}) |}} |U_{j_1},
  \ldots, U_{j_{| \mathcal{D} (\ensuremath{\boldsymbol{x}}) | - 1}},
  \ensuremath{\boldsymbol{X}}$
\end{itemize}

\textcolor{blue}{We now note that the order of the indices $j_1,...,j_{| \mathcal{D} (\ensuremath{\boldsymbol{x}}) |}$ is irrelevant for the sampling scheme. Although it might appear that the sampling procedure depends on the ordering of those indices, the acceptance or rejection of such samples also depends on the ordering and the next subsection shows
that such a procedure will always result in a correct MCMC draw from the conditional distribution $\ensuremath{\boldsymbol{U}}_{\mathcal{D}
(\ensuremath{\boldsymbol{x}})} |\ensuremath{\boldsymbol{X}}$.}

The above sampling scheme requires knowing the marginal distribution of
$\ensuremath{\boldsymbol{U}}_{\mathcal{J}} |\ensuremath{\boldsymbol{X}}$ for
$\mathcal{J} \subset \mathcal{D} (\ensuremath{\boldsymbol{x}})$ and the
conditional decomposition $U_j |\ensuremath{\boldsymbol{U}}_{\mathcal{K}},
\ensuremath{\boldsymbol{X}}$ where $ (\{ j \}, \mathcal{K})$ is a partition of
$\mathcal{J}$ (meaning $\{ j \} =\mathcal{J} \backslash \mathcal{K}$, the
complement of $\mathcal{K}$ in $\mathcal{J}$). This distribution can be
derived as
\begin{eqnarray*}
  f (\ensuremath{\boldsymbol{u}}_{\mathcal{J}} |\ensuremath{\boldsymbol{x}}) &
  = & \frac{c (\ensuremath{\boldsymbol{b}}_{\mathcal{C}
  (\ensuremath{\boldsymbol{x}})}) \prod_{j \in \mathcal{C}
  (\ensuremath{\boldsymbol{x}})} f (x_j)}{f (\ensuremath{\boldsymbol{x}})} c
  (\ensuremath{\boldsymbol{u}}_{\mathcal{J}}
  |\ensuremath{\boldsymbol{b}}_{\mathcal{C} (\ensuremath{\boldsymbol{x}})})\\
  & \times & \left[
  \bigtriangleup_{\ensuremath{\boldsymbol{a}}_{\mathcal{J}^c}}^{\ensuremath{\boldsymbol{b}}_{\mathcal{J}^c}}
  C_{\ensuremath{\boldsymbol{U}}_{\mathcal{J}^c}
  |\ensuremath{\boldsymbol{U}}_{\mathcal{J}},
  \ensuremath{\boldsymbol{U}}_{\mathcal{C} (\ensuremath{\boldsymbol{x}})}}
  (\cdot |\ensuremath{\boldsymbol{u}}_{\mathcal{J}},
  \ensuremath{\boldsymbol{b}}_{\mathcal{C} (\ensuremath{\boldsymbol{x}})})
  \right] \prod_{j \in \mathcal{J}} \mathcal{I} (a_j \leqslant u_j < b_j)
\end{eqnarray*}
with  $\mathcal{J}^c =\mathcal{D} (\ensuremath{\boldsymbol{x}}) \backslash
\mathcal{J}$ and
\begin{eqnarray*}
  f (u_j |\ensuremath{\boldsymbol{u}}_{\mathcal{K}},
  \ensuremath{\boldsymbol{x}}) & = & c (u_j
  |\ensuremath{\boldsymbol{u}}_{\mathcal{K}},
  \ensuremath{\boldsymbol{b}}_{\mathcal{C} (\ensuremath{\boldsymbol{x}})})
  \mathcal{I} (a_j \leqslant u_j < b_j)\\
  & \times &
  \frac{\bigtriangleup_{\ensuremath{\boldsymbol{a}}_{\mathcal{J}^c}}^{\ensuremath{\boldsymbol{b}}_{\mathcal{J}^c}}
  C_{\ensuremath{\boldsymbol{U}}_{\mathcal{J}^c}
  |\ensuremath{\boldsymbol{U}}_{\mathcal{J}},
  \ensuremath{\boldsymbol{U}}_{\mathcal{C}}} (\cdot
  |\ensuremath{\boldsymbol{u}}_{\mathcal{J}},
  \ensuremath{\boldsymbol{b}}_{\mathcal{C}
  (\ensuremath{\boldsymbol{x}})})}{\bigtriangleup_{\ensuremath{\boldsymbol{a}}_{\mathcal{K}^c}}^{\ensuremath{\boldsymbol{b}}_{\mathcal{K}^c}}
  C_{\ensuremath{\boldsymbol{U}}_{\mathcal{K}^c}
  |\ensuremath{\boldsymbol{U}}_{\mathcal{K}},
  \ensuremath{\boldsymbol{U}}_{\mathcal{C}}} (\cdot
  |\ensuremath{\boldsymbol{u}}_{\mathcal{K}},
  \ensuremath{\boldsymbol{b}}_{\mathcal{C} (\ensuremath{\boldsymbol{x}})})},
\end{eqnarray*}
where $\mathcal{K}^c =\mathcal{J}^c \cup \{ j \}$.

\textcolor{blue}{We continue to illustrate how to apply the latent variables conditional formulas by considering Example~\ref{ex: example 1}.}
\setcounter{example}{0}
\begin{example}[continued]
If $x_1 \neq 0$, then
\[ f (u_2 |\tmmathbf{x}) = \frac{c_{2|1} (u_2 |F_1 (x_1)) \mathcal{I} (F_2
   (x_2 -) \leqslant u_2 < F_2 (x_2))}{C_{2|1} (F_2 (x_2) |F_1 (x_1)) -
   C_{2|1} (F_2 (x_2 -) |F_1 (x_1))} \]
($u_1$ is deterministically equal to $F_1 (x_1)$, so we only need to sample
$u_2$).

If $x_1 = 0$
\[ f (u_1, u_2 |\tmmathbf{x}) = \frac{c (u_1, u_2) \mathcal{I} (F_1 (0 -)
   \leqslant u_1 < F_1 (0)) \mathcal{I} (F_2 (x_2 -) \leqslant u_2 < F_2
   (x_2))}{C (F_1 (0), F_2 (x_2)) - C (F_1 (0), F_2 (x_2 -)) - C (F_1 (0 -),
   F_2 (x_2)) + C (F_1 (0 -), F_2 (x_2 -))} \]

\end{example}

\subsection{Metropolis-Hastings  sampling}
It is clear from the formulas for $f (u_j
|\ensuremath{\boldsymbol{u}}_{\mathcal{K}}, \ensuremath{\boldsymbol{x}})$
that they are quite intricate. They correspond to a product of a simple term
$c (u_j |\ensuremath{\boldsymbol{u}}_{\mathcal{K}},
\ensuremath{\boldsymbol{b}}_{\mathcal{C} (\ensuremath{\boldsymbol{x}})})
\mathcal{I} (a_j \leqslant u_j < b_j)$ (a truncated conditional copula density)
and a complicated term that depends on ratios of normalizing constants for
$f (\ensuremath{\boldsymbol{u}}_{\mathcal{J}} |\ensuremath{\boldsymbol{x}})$
and $f (\ensuremath{\boldsymbol{u}}_{\mathcal{K}}
|\ensuremath{\boldsymbol{x}})$. One of the most useful aspects of the
Metropolis-Hastings (MH) algorithm is that it does not require knowledge of normalizing
constants. The trick here is that those normalizing constants are obtained
recursively. Assume that we sample
\begin{itemize}
  \item $U_{j_1}$ from $c (u_{j_1}) \mathcal{I} (a_{j_1} \leqslant u_{j_1} <
  b_{j_1})$

  \item $U_{j_2}$ from $c (u_{j_2} |u_{j_1}) \mathcal{I} (a_{j_2} \leqslant
  u_{j_2} < b_{j_2})$

  \item $\vdots$

  \item $U_{j_{| \mathcal{D} (\ensuremath{\boldsymbol{x}}) |}}$ from $c
  (u_{j_{| \mathcal{D} (\ensuremath{\boldsymbol{x}}) |}} |u_{j_1}, \ldots,
  u_{j_{| \mathcal{D} (\ensuremath{\boldsymbol{x}}) | - 1}}) \mathcal{I}
  (a_{j_{| \mathcal{D} (\ensuremath{\boldsymbol{x}}) |}} \leqslant u_{j_{|
  \mathcal{D} (\ensuremath{\boldsymbol{x}}) |}} < b_{j_{| \mathcal{D}
  (\ensuremath{\boldsymbol{x}}) |}})$
\end{itemize}
that is, if we use as proposal a truncated form of the copula marginal density
over $\mathcal{D} (\ensuremath{\boldsymbol{x}})$, then computing the MH
accept/reject ratio results in the computationally simple formula
\[ \alpha (\ensuremath{\boldsymbol{x}}_i) = \prod_{k = 1}^{| \mathcal{D}
   (\ensuremath{\boldsymbol{x}}) |} \frac{C (F_{j_k} (x_{i, j_k}) |u^N_{i,
   j_1}, \ldots, u^N_{i, j_{k - 1}}, \ensuremath{\boldsymbol{b}}_{\mathcal{C}
   (\ensuremath{\boldsymbol{x}}_i), i}) - C (F_{j_k} (x_{i, j_k}^-) |u^N_{i,
   j_1}, \ldots, u^N_{i, j_{k - 1}}, \ensuremath{\boldsymbol{b}}_{\mathcal{C}
   (\ensuremath{\boldsymbol{x}}_i), i})}{C (F_{j_k} (x_{i, j_k}) |u^O_{i,
   j_1}, \ldots, u^O_{i, j_{k - 1}}, \ensuremath{\boldsymbol{b}}_{\mathcal{C}
   (\ensuremath{\boldsymbol{x}}_i), i}) - C (F_{j_k} (x_{i, j_k}^-) |u^O_{i,
   j_1}, \ldots, u^O_{i, j_{k - 1}}, \ensuremath{\boldsymbol{b}}_{\mathcal{C}
   (\ensuremath{\boldsymbol{x}}_i), i})} \]
where $i$ represents the observation index. The complexity of this formula is
much smaller than $2^{| \mathcal{D} (\ensuremath{\boldsymbol{x}}) |}$.

\textcolor{blue}{We now illustrate the Metropolis-Hastings acceptance probabilities by again considering Example~\ref{ex: example 1}.}
\setcounter{example}{0}
\begin{example}[continued]
If $x_1 \neq 0$, then the ratio is
$ \alpha (x_2) = 1 $
and if $x_1 = 0$ (first draw $u_1^N$ from a uniform on $(F_1 (0^-), F_1 (0))$
and compare to the previous draw $u_1^O$)
\[ \alpha (0, x_2) = \frac{C_{2|1} (F_2 (x_2) |u_1^N) - C_{2|1} (F_2 (x_2^-)
   |u_1^N)}{C_{2|1} (F_2 (x_2) |u_1^O) - C_{2|1} (F_2 (x_2^-) |u_1^O)} \]
Note that here the ordering does not matter, as we could have computed the
other ratio (if we draw instead first $u_2^N$ from a uniform on $(F_2 (x_2^-),
F_2 (x_2))$
\[ \alpha (0, x_2) = \frac{C_{1|2} (F_1 (0) |u_2^N) - C_{1|2} (F_1 (0^-)
   |u_2^N)}{C_{1|2} (F_1 (0) |u_2^O) - C_{1|2} (F_1 (0^-) |u_2^O)} \]
Even though the ratio are different, both procedures will result in a draw
from $f (u_1, u_2 |\tmmathbf{x})$.

\end{example}

\subsection{Mixtures of Archimedean and Gaussian copulas\label{SS: mixtures of archimedean and gaussian copulas}}
This section applies the previous results to the family of mixtures of Archimedean and Gaussian copulas. Working with mixtures of copulas
provides a simple and yet rich and flexible modeling framework because mixtures of copulas are copulas themselves,

We are particularly interested in having a mixture of three components, two Archimedean copulas, the Clayton copula
$\left(C_{Cl}\right)$ and the Gumbel copula $\left(c_{Gu}\right)$ and a Gaussian copula $\left(c_{G}\right)$
component. We will later apply this mixture to model the dependence between individual income distributions
over 13 years. The copula density of this 3-component mixture is
\begin{equation}
c_{mix}\left(\boldsymbol{u};\Gamma,\theta_{Cl},\theta_{Gu},w_{1},w_{2}\right)=w_{1}c_{G}\left(\boldsymbol{u};\Gamma\right)+w_{2}c_{Cl}\left(\boldsymbol{u};\theta_{Cl}\right)
+w_{3}c_{Gu}\left(\boldsymbol{u};\theta_{Gu}\right), \label{eq:mixture copula model}
\end{equation}
where $w_{1}$, $w_{2}$, and $w_{3}=1-w_{1}-w_{2}$ are the mixture
weights, and $\Gamma$, $\theta_{Cl}$, and $\theta_{Gu}$ are respectively the
dependence parameters of the Gaussian, Clayton, and Gumbel copulas.
Such a mixture of copula models has the additional flexibility
of being to capture lower and upper tail dependence. We will use a Bayesian approach
to estimate the copula parameters and, for simplicity and without loss of generality, we follow \cite{joe2014dependence} and use empirical CDF's
to model the marginal distributions.

Let the parameter $w_{k}$ denote the probability that the $i$-th
observation comes from the $k$-th component in the mixture. Let $\boldsymbol{d}_{i}=\left(d_{i1},d_{i2},d_{i3}\right)^{'}$
be indicator (latent) variables such that $d_{ik}=1$ when the $i$-th
observation comes from the $k$-th component in the mixture. These
indicator variables identify the component of the copula model defined
in equation \eqref{eq:mixture copula model} to which the observation
$\boldsymbol{y}_{i}$ belongs. Then,

\begin{equation}
p\left(d_{ik}=1|\boldsymbol{w}\right)=w{}_{k},
\end{equation}
with $w_{k}>0$ and $\sum_{k=1}^{3}w_{k}=1$.

Given the information on the $n$ independent sample observations
$\boldsymbol{y}=\left(\boldsymbol{y}_{1},...,\boldsymbol{y}_{n}\right)^{'}$
and $\boldsymbol{y}_{i}=\left(y_{i1},...,y_{iT}\right)$, and by using
Bayes rule, the joint posterior density is  obtained as
\begin{equation}
p\left(\boldsymbol{w},\boldsymbol{d},\Gamma,\theta_{Cl},\theta_{Gu}|\boldsymbol{y}\right)\propto p\left(\boldsymbol{y}|\boldsymbol{w},\boldsymbol{d},\Gamma,\theta_{Cl},\theta_{Gu}\right)
p\left(\boldsymbol{d}|\boldsymbol{w},\Gamma,\theta_{Cl},\theta_{Gu}\right)p\left(\boldsymbol{w}\right)p\left(\Gamma\right)p\left(\theta_{Cl}\right)
p\left(\theta_{Gu}\right)\label{eq:posterior mixture copula}
\end{equation}
with
\[
p\left(\boldsymbol{y}|\boldsymbol{w},\boldsymbol{d},\Gamma,\theta_{Cl},\theta_{Gu}\right)=\prod_{i=1}^{n}\left[c_{G}\left(\boldsymbol{u};\Gamma\right)\right]^{d_{i1}}
\left[c_{Cl}\left(\boldsymbol{u};\theta_{Cl}\right)\right]^{d_{i2}}\left[c_{Gu}\left(\boldsymbol{u};\theta_{Gu}\right)\right]^{d_{i3}},
\]
and
\begin{equation}
p\left(\boldsymbol{d}|\boldsymbol{w},\Gamma,\theta_{Cl},\theta_{Gu}\right)=p\left(\mathbf{d}|\boldsymbol{w}\right)=
\prod_{i=1}^{n}\prod_{k=1}^{K}w_{k}^{d_{ik}}=\prod_{k=1}^{K}w_{k}^{n_{k}},
\end{equation}
where $n_{k}=\sum_{i=1}^{n}I\left(d_{ik}=1\right)$ and $I\left(d_{ik}=1\right)$
is an indicator variable which is equal 1 if observation
$i$ belongs to the $k$-th component of the copula mixture model,
and is $0$ otherwise. We use a Dirichlet prior for $\boldsymbol{w}$,
$p\left(\boldsymbol{w}\right)=Dirichlet\left(\boldsymbol{\phi}\right)$, which is
defined as
\begin{equation}
p\left(\boldsymbol{w}\right)\propto w_{1}^{\phi_{1}-1}...w_{3}^{\phi_{3}-1}. \label{eq:weight prior}
\end{equation}
The Dirichlet distribution is the common choice in Bayesian mixture
modeling since it is a conjugate of the multinomial distribution
\citep{Diebold1994} . We use the gamma density $G(\alpha,\beta)$ as the prior distribution for $\theta_{Cl}$ and $\theta_{Gu}$.
The hyperparameters
in the prior PDFs are chosen so that the priors are uninformative.
We use a Metropolis within Gibbs sampling algorithm to draw observations
from the joint posterior PDF defined in equation~\eqref{eq:posterior mixture copula}
and use the resulting MCMC draws to estimate the quantities required for inference.
The relevant conditional posterior PDFs are now specified.

The conditional posterior probability that the $i$th observation
comes from the $k$th component in the copula mixture model is
\begin{equation}
p\left(d_{ik}|\boldsymbol{w},\Gamma,\theta_{Cl},\theta_{Gu},\boldsymbol{y}\right)=\frac{p_{ik}}{p_{i1}+...+p_{i3}}, \label{eq:eq 17}
\end{equation}
where $p_{i1}=w_{1}c_{G}\left(\boldsymbol{u};\Gamma\right)$, $p_{i2}=w_{2}c_{Cl}\left(\boldsymbol{u};\theta_{Cl}\right)$,
and $p_{i3}=w_{3}c_{Gu}\left(\boldsymbol{u};\theta_{Gu}\right)$ for
$k=1,2,3$.
The conditional posterior PDF for the mixture weights $\boldsymbol{w}$
is the Dirichlet PDF
\begin{equation}
p(\boldsymbol{w}|\mathbf{d},\Gamma,\theta_{Cl},\theta_{Gu},\boldsymbol{y})=D(\boldsymbol{\phi}+\boldsymbol{n}), \label{eq:eq 28}
\end{equation}
where $\mathbf{n}=(n_{1},...,n_{k})'$ and $\boldsymbol{\phi}=(\phi_{1},...,\phi_{K})'$.
The conditional posterior PDF for the Gaussian copula parameter matrix $\Gamma$
is
\begin{equation}
p\left(\Gamma|\boldsymbol{y},\mathbf{d},\theta_{Cl},\theta_{Gu},\boldsymbol{w}\right)=\prod_{i\in d_{i1}=1}c_{G}\left(\boldsymbol{u};\Gamma\right)p\left(\Gamma\right). \label{eq:gaussian conditiona}
\end{equation}
The conditional posterior PDF for the Clayton copula parameter $\theta_{Cl}$
is
\begin{equation}
p\left(\theta_{Cl}|\boldsymbol{y},\mathbf{d},\Gamma,\theta_{Gu},\boldsymbol{w}\right)=\prod_{i\in d_{i1}=2}c_{Cl}\left(\boldsymbol{u};\theta_{Cl}\right)p\left(\theta_{Cl}\right). \label{eq:clayton conditional}
\end{equation}
The conditional posterior PDF  for the Gumbel copula parameter $\theta_{Gu}$
is
\begin{equation}
p\left(\theta_{Gu}|\boldsymbol{y},\mathbf{d},\Gamma,\theta_{Cl},\boldsymbol{w}\right)=\prod_{i\in d_{i1}=3}c_{Gu}\left(\boldsymbol{u};\theta_{Gu}\right)p\left(\theta_{Gu}\right). \label{eq:gumbel conditional}
\end{equation}
Generating the conditional posterior density for $\theta_{Cl}$ and
$\theta_{Gu}$ is not straightforward since the conditional posterior
densities for both $\theta_{Cl}$ and $\theta_{Gu}$ are not in a recognizable
form. We use a random walk Metropolis algorithm to draw
from the conditional posterior densities of both $\theta_{Cl}$
and $\theta_{Gu}$. The generation of the Gaussian copula matrix parameter $\Gamma$
is more complicated and is explained in the next section.

The full MCMC sampling scheme is
\begin{enumerate}
\item Set the starting values for $\boldsymbol{w}^{\left(0\right)}$, $\Gamma^{\left(0\right)}$,
$\theta_{Cl}^{\left(0\right)}$, and $\theta_{Gu}^{\left(0\right)}$
\item Generate $(\boldsymbol{w}^{\left(t+1\right)}|\mathbf{d}^{\left(t\right)},\Gamma^{\left(t\right)},\theta_{Cl}^{\left(t\right)},\theta_{Gu}^{\left(t\right)},\boldsymbol{y})$
from equation \eqref{eq:eq 28}
\item Generate $\left(\Gamma^{\left(t+1\right)}|\boldsymbol{y},\mathbf{d}^{\left(t+1\right)},\theta_{Cl}^{\left(t\right)},\theta_{Gu}^{\left(t\right)},\boldsymbol{w}^{\left(t+1\right)}\right)$
from equation \eqref{eq:gaussian conditiona}
\item Generate $\left(\theta_{Cl}^{\left(t+1\right)}|\boldsymbol{y},\mathbf{d}^{\left(t+1\right)},\Gamma^{\left(t+1\right)},\theta_{Gu}^{\left(t\right)},\boldsymbol{w}^{\left(t+1\right)}\right)$
from equation \eqref{eq:clayton conditional}
\item Generate $\left(\theta_{Gu}^{\left(t+1\right)}|\boldsymbol{y},\mathbf{d}^{\left(t+1\right)},\Gamma^{\left(t+1\right)},\theta_{Cl}^{\left(t+1\right)},\boldsymbol{w}^{\left(t+1\right)}\right)$
from equation \eqref{eq:gumbel conditional}
\item Set $t=t+1$ and return to step 2.
\end{enumerate}

Appendix~\ref{app_sampling_schemes}  gives further details on the particulars of the sampling scheme. In particular, it describes how to write the distributions and densities of the Gaussian, Gumbel and Clayton copulas respectively and how to sample from them. It also details how to sample the correlation parameters of the Gaussian copula and summarizes how the one-margin at a time latent variable simulation works.

\section{Application to Individual Income Dynamics\label{S: application to individual income dynamics}}

Longitudinal or panel datasets, such as the Panel Study of Income Dynamics
(PSID), the British Household Panel Survey (BHPS), and the Household Income and
Labour Dynamics Survey in Australia (HILDA) are increasingly used for
assessing income inequality, mobility, and poverty over time.
The income data from these surveys for different years are correlated
due to the nature of panel studies. For such correlated samples, the
standard approach of fitting univariate models to income distributions
for different years may give rise to misleading results. The univariate
approach treats the income distribution over different years as independent
and ignores the dependence structure between incomes for different years. It does not
take into account that those who earned a high income in one year are more likely to earn
a high income in subsequent years and vice versa. A
common way to address this problem is to use a multivariate income
distribution model that takes into account the dependence between
incomes for different years.

The presence of dependence in a sample of incomes from panel datasets
has rarely been addressed in the past. Only recently, \cite{Vinh2010}
proposed using bivariate copulas to model income distributions for
two different years,  using maximum likelihood estimation. However, in their applications, they do not take
into account the point mass occurring at zero income. Our methodology
is more general than \cite{Vinh2010}. We estimate a panel of incomes
from the HILDA survey from 2001 to 2013 using a finite mixture of Gaussian, Clayton,
and Gumbel copulas while taking into account the point mass occurring
at zero incomes. Once the parameters for the multivariate income distribution
have been estimated, values for various measures of inequality, mobility,
and poverty can be obtained. Our methodology is Bayesian which enables us to estimate
the  posterior densities of  the
parameters of the copula models and the inequality, mobility,
and poverty measures. In this example, we consider the \cite{Shorrocks1978a}
and \cite{Foster2009} indices for illustration purposes. Other inequality,
mobility, and poverty indices can be estimated similarly.
For other recent studies on income mobility dynamics, see also \cite{Bonhomme2009}.

Although a number of income related variables are available, we use
the imputed income series \_WSCEI in this example. This variable
contains the average individual weekly wage and salary incomes from
all paid employment over the period considered. It is reported before
taxation and governmental transfers. The income data were also adjusted
to account for the effects of inflation using the Consumer Price Index
data obtained from the Australian Bureau of Statistics, which is based
in 2010 dollars. From these data, a dependence sample was constructed
by establishing whether a particular individual had recorded an income
in all the years. Individuals who only recorded incomes in some
of the years being considered were removed. In addition, we also focus
our attention on individuals who are in the labor force (both employed
and unemployed). We found that 1745 individuals recorded an income
for all 13 years. Table~\ref{tab:Descriptive-statistics-for}
summarizes the distributions of real individual disposable income in Australia for the years 2001
- 2013 and shows that all income distributions
exhibit positive skewness and fat long right tails typical of
income distributions. If  the ordering of the
distributions is judged on the basis of the means or the medians, the population
becomes better off as it moves from 2001 to 2013, except between the period
2006 and 2007.  These effects are also confirmed by Figures~\ref{fig:Histogram-of-real} to
\ref{fig:Histogram-of-real-2} in appendix~\ref{app: extra empirical results}

\begin{sidewaystable}[]

\centering{}\caption{Descriptive statistics for real individual wages for Australia for
the years 2001 - 2013\label{tab:Descriptive-statistics-for}}
\begin{tabular}{cccccccccccccc}
\hline
 & 2001 & 2002 & 2003 & 2004 & 2005 & 2006 & 2007 & 2008 & 2009 & 2010 & 2011 & 2012 & 2013\tabularnewline
\hline
\hline
Mean & 684 & 734 & 766 & 819 & 874 & 923 & 783 & 1067 & 1105 & 1128 & 1188 & 1215 & 1245\tabularnewline
Median & 616 & 673 & 712 & 753 & 803 & 852 & 702& 969 & 1003 & 1048 & 1051 & 1101 & 1100 \tabularnewline
Std. dev. & 551 & 591 & 568 & 645 & 674 & 668 & 694 & 788 & 825 & 869 & 990 & 916 & 950 \tabularnewline
skewness & 2.1 & 3.0 & 2.1 & 3.7 & 2.9 & 1.5 & 2.0& 2.0 & 1.8 & 2.0 & 3.7 & 1.9 & 1.5 \tabularnewline
kurtosis & 15.7 & 26.6 & 16.0 & 40.2 & 27.1 & 8.7 & 11.1& 12.7 & 11.7 & 15.7 & 37.5 & 13.0 & 7.4 \tabularnewline
\hline
\end{tabular}
\end{sidewaystable}

\subsection{Foster's (2009) Chronic Poverty Measures}

The measurement of chronic income poverty is important because it focuses
on those whose lack of income stops them from obtaining the ``minimum necessities
of life'' for much of their life course. Let $z\in \mathbb{R^{+}}$ be the
poverty line. It is the level of income/wages which is just sufficient for someone
to be able to afford the minimum necessities of life. For every $i=1,...,n$
and $t=1,...,T$, the row vector $\mathbf{y}_{i}=\left(y_{i1},...,y_{iT}\right)$
contains individual $i$'s incomes across time and the column vector
$\mathbf{y}_{*t}=\left(y_{1t},...,y_{nt}\right)^{'}$ contains the
income distribution at period $t$.

The measurement of chronic poverty is split into two steps:
an ``identification'' step and an aggregation step. The identification
function $\rho\left(\mathbf{y}_{i};z\right)$ indicates that individual
$i$ is in chronic poverty when $\rho\left(\mathbf{y}_{i};z\right)=1$,
while $\rho\left(\mathbf{y}_{i};z\right)=0$ otherwise. \cite{Foster2009} proposed an identification
method that counts the number of periods of poverty experienced by a particular
individual, $y_{it}<z$, and then expressed it as a fraction $d_{i}$
of the $T$ periods. The identification function $\rho_{\tau}\left(\mathbf{y}_{i};z\right)=1$
if $d_{i}\geq\tau$ and $\rho_{\tau}\left(\mathbf{y}_{i};z\right)=0$
if $d_{i}<\tau$.

The aggregation step combines the information on the chronically poor
people to obtain an overall level of chronic poverty in a given society.
We use the extension of univariate Foster, Greer and Thorbecke (FGT) poverty indices
of \citet{Foster:1984}.
These are given by
\[
FGT^{\alpha}\left(z\right)=\frac{1}{n}\sum_{i=1}^{n}g_{i}^{\alpha}\ ,
\]
where $g_{i}^{\alpha}=0$ if $y_{i}>z$ and $g_{i}^{\alpha}\left(z\right)=\left(\frac{z-y_{i}}{z}\right)^{\alpha}$
if $y_{i}\leq z$, and $\alpha$ measures inequality aversion. The
FGT measure when $\alpha=0$ is called the headcount ratio,
when $\alpha=1$ it is called the poverty gap index and when
$\alpha=2$ it is called the poverty severity index. \cite{Foster2009} proposed duration
adjusted FGT poverty indices: duration adjusted headcount ratio and duration
adjusted poverty gap. Following \cite{Foster2009}, we define the normalized
gap matrix as $G^{\alpha}\left(z\right):=\left[g_{it}^{\alpha}\left(z\right)\right]$,
where $g_{it}^{\alpha}\left(z\right)=0$ if $y_{it}>z$ and $g_{it}^{\alpha}\left(z\right)=\left(\frac{z-y_{it}}{z}\right)^{\alpha}$
if $y_{it}\leq z$. Then, identification is incorporated into the
censored matrix $G^{\alpha}\left(z,\tau\right)=\left[g_{it}^{\alpha}\left(z,\tau\right)\right]$,
where $g_{it}^{\alpha}\left(z,\tau\right)=g_{it}^{\alpha}\left(z\right)\rho_{\tau}\left(\mathbf{y}_{i};z\right)$.
The entries for the non-chronically poor are censored to zero, while the
entries for the chronically poor are left unchanged. When $\alpha=0$,
the measure becomes the duration adjusted headcount ratio and is
the mean of $G^{0}\left(z,\tau\right)$, and when $\alpha=1$,
the measure becomes the duration adjusted poverty gap, and is given
by the mean of $G^{1}\left(z,\tau\right)$.

\subsection{Shorrocks (1978a) Income Mobility Measures\label{SS: income mobility measures}}

The measurement of income mobility focuses on how individuals' income
changes over time. Many mobility measures have been developed and applied
to empirical data to describe income dynamics; see \cite{Shorrocks1978a},
\cite{Shorrocks1978b}, \cite{Formby2004}, \cite{Dardanoni1993},
\cite{Fields1996}, \cite{Maasoumi1986}, and references therein.
However, statistical inference on income mobility has been largely
neglected in the literature. Only recently, some researchers have
developed statistical inference procedures for the measurement of
income mobility \citep{Biewen2002, Maasoumi2001, Formby2004}.
Here, we show that our approach can be used to obtain the posterior densities
of mobility measures which can then be used for making inference on income mobility.

\cite{Shorrocks1978a} proposed a measure of income mobility that
is based on transition matrices. Following \cite{Formby2004}, we
consider the joint distribution of two income variables $y_{1}$
and $y_{2}$ with a continuous CDF $F\left(y_{1},y_{2}\right)$. This
distribution function captures all the transitions between $y_{1}$ and $y_{2}$.
In this application, we consider the mobility between two points in
time. The movement between $y_{1}$ and $y_{2}$ is described by a
transition matrix. To form the the transition matrix from $F\left(y_{1},y_{2}\right)$,
we need to determine the number of, and boundaries between,  income classes.
Suppose there are $m$ classes in each of the income variables
and the boundaries of these classes are $0<\tau_{1}^{y_{1}}<...<\tau_{m-1}^{y_{1}}<\infty$
and $0<\tau_{1}^{y_{2}}<...<\tau_{m-1}^{y_{2}}<\infty$. The resulting
transition matrix is denoted $P=\left[p_{ij}\right]$. Each element
$p_{ij}$ is a conditional probability that an individual moves to
class $j$ of income $y_{2}$ given that they are initially in class
$i$ with income $y_{1}$. It is defined as
\[
p_{ij}:=\frac{\Pr\left(\tau_{i-1}^{y_{1}}\leq y_{1}<\tau_{i}^{y_{1}}\;and\;\tau_{j-1}^{y_{2}}\leq y_{2}<\tau_{j}^{y_{2}}\right)}{\Pr\left(\tau_{i-1}^{y_{1}}\leq y_{1}<\tau_{i}^{y_{1}}\right)}\ ,
\]
where $\Pr\left(\tau_{i-1}^{y_{1}}\leq y_{1}<\tau_{i}^{y_{1}}\right)$
is the probability that an individual falls into income class $i$
of $y_{1}$.

A Mobility measure $M\left(P\right)$ can be defined as a function
of the transition matrix $P$. We say that a society with transition matrix
$P_{1}$ is more mobile than one with transition matrix $P_{2}$, according
to mobility measure $M\left(P\right)$, if and only if $M\left(P_{1}\right)>M\left(P_{2}\right)$.
We consider a mobility measure developed by \cite{Shorrocks1978a} and defined as
 \[
M_{1}\left(P\right):=\frac{m-\sum_{i=1}^{m}p_{ii}}{m-1};
\]
$M_{1}$ measures the average probability across all classes that
an individual will leave his initial class in the next period.

\subsection{Empirical Analysis\label{SS: empirical analysis}}

This section discusses  the results from the analysis
of the real individual wages data after estimating the proposed multivariate
income distribution model using a Bayesian approach.
The univariate income distribution is usually modeled using  Dagum or Singh-Maddala distributions \citep{Kleiber1996}.
In this example, the marginal income distribution is modeled using empirical distribution function,
for simplicity. It is straightforward to extend the MCMC sampling scheme in Section~\ref{S: estimation and algorithms}
to estimate both marginal
and joint parameters as in  \cite{Pitt2006} and \cite{Smith2012}.

First, we present
the model selection results and the estimated parameters of the copula models.
To select the best copula model, we use the $DIC_{3}$ criterion of \cite{Celeux:2006}
and the cross-validated log predictive score (LPDS) \citep{Good:1952, Geisser1980}. The $DIC_{3}$ criterion is defined as
\[
DIC_{3}:=-4\E_{\theta}\left(\log p\left(\boldsymbol{y}|\theta\right)|\boldsymbol{y}\right)+2\log\widehat{p}\left(\boldsymbol{y}\right),
\]
where $\widehat{p}\left(\boldsymbol{y}\right)=\prod_{i=1}^{n}\widehat{p}\left(y_{i}\right)$.
We next define the $B$-fold cross-validated LPDS.
Suppose that the dataset $\D$ is split into roughly $B$ equal
parts $\D_{1},...,\D_{B}$. Then, the $B-$fold cross validated LPDS is
defined as
\[
LPDS\left(\widehat{p}\right):=\sum_{j=1}^{B}\sum_{y_{j}\in\D_{j}}\log\widehat{p}\left(\mathbf{y}_{j}|\D\setminus\D_{j}\right)
\]
In our work we take $B = 5 $.
Table \ref{tab:Model-Selection-} shows that the best model, according to both criteria, is the
mixture of Gaussian, Clayton, and Gumbel copulas.
We estimate the best model with an initial burnin period of 10000
sweeps and a Monte Carlo sample of 10000 iterates. Next, we use the iterates from the best
model to estimate transition probabilities from 0 to positive
wages and from positive wages to zero, Spearman's correlation coefficient, and the mobility
and poverty measures, by averaging over the posterior distribution of the parameters.

Table~\ref{tab:Some-Estimated-Parameters copula} shows some of the estimated
parameters and corresponding 95\% credible intervals for the chosen
copula mixture model. The parameters and their 95\% credible intervals
are quite tight, indicating that the parameters are well estimated.
It is clear that there are significant differences in the estimated parameters
by taking into account the point mass at zero wages compared to the parameters
estimated by not taking into account this point mass. The estimated
mixture weight parameters show that the Gaussian copula has the
highest weight, followed by the Clayton and Gumbel copulas. As the weight
of the Clayton copula is higher than of the Gumbel copula, it implies that
there are more people with lower tail dependence than upper tail dependence.
This may coincide with a relatively higher degree of income
mobility amongst high income earners.

\begin{table}[H]
\caption{Model Selection of the copula to model 13 years of income distribution with point mass at zero incomes \label{tab:Model-Selection-} }

\centering{}%
\begin{tabular}{ccc}
\hline
Model & $DIC_{3}$ & LPDS-CV\tabularnewline
\hline
Clayton & $-1.21\times10^{4}$ & $6.03\times10^{3}$\tabularnewline
Gumbel & $-1.75\times10^{4}$ & $4.95\times10^{3}$\tabularnewline
Gaussian  & $-2.13\times10^{4}$ & $4.29\times10^{4}$\tabularnewline
Mixture (Gaussian, Clayton) & $-2.86\times10^{4}$ & $5.63\times10^{4}$\tabularnewline
Mixture (Gaussian, Gumbel) & $-2.83\times10^{4}$ & $5.54\times10^{4}$\tabularnewline
Mixture (Clayton, Gumbel) & $-1.68\times10^{4}$ & $3.31\times10^{4}$\tabularnewline
Mixture (Gaussian, Clayton, Gumbel) & $-2.89\times10^{4\star}$ & $5.68\times10^{4\star}$\tabularnewline
\hline
\end{tabular}
\end{table}


\begin{table}[H]
\caption{Some of the estimated parameters of the mixture of the Gaussian, Gumbel and Clayton copulas to model 13 years of income distributions. The 95\% credible intervals are in brackets
 \label{tab:Some-Estimated-Parameters copula}}
\centering{}%
\begin{tabular}{ccc}
\hline
Parameters & \multicolumn{1}{c}{Copula (Point Mass)} & \multicolumn{1}{c}{Copula (No Point Mass)}\tabularnewline
\hline
\hline
$\theta_{Cl}$ & $\underset{\left(0.12,0.18\right)}{0.15}$ & $\underset{\left(0.29,0.37\right)}{0.33}$\tabularnewline
$\theta_{Gu}$ & $\underset{\left(1.84,2.06\right)}{1.94}$ & $\underset{\left(2.23,2.45\right)}{2.33}$\tabularnewline
$w_{1}$ & $\underset{\left(0.64,0.69\right)}{0.66}$ & $\underset{\left(0.60,0.65\right)}{0.62}$\tabularnewline
$w_{2}$ & $\underset{\left(0.19,0.24\right)}{0.21}$ & $\underset{\left(0.21,0.26\right)}{0.23}$\tabularnewline
\hline
\end{tabular}
\end{table}
Tables~\ref{tab:Transition-Probability aug-2} and
\ref{tab:Transition-Probability aug-1-1} in Appendix~\ref{app: extra empirical results}
 present the estimates
of the transition probabilities from 0 to positive wages and from positive
to 0 wages. The estimates of the transition probabilities seem to be
close to their sample (non-parametric) counterparts. The estimates of
transition probabilities from 0 to positive wages are similar
(0.39-0.49) in the period from 2001-2006. Similarly, the
estimates are similar  in the period 2008-2013 (0.34-0.38).
However,  there are higher estimates for the period 2006-2007 and 2007-2008 (0.83 and 0.87,
respectively). Similar
results are observed for the transition probabilities from positive to zero wages.
 The estimates of the transition probabilities are
roughly the same between the periods 2001-2006 and 2008-2013. There
are higher estimates for the period 2006-2007 and 2007-2008. This
phenomenon may indicate that there is very high income mobility between
2006-2007 and 2007-2008. Note that the model that does not take into account the point masses at zero cannot give us the estimate of transition probabilities.

\begin{figure}[H]

\caption{Estimated headcount posterior densities based on including (left density-blue line) and not including point masses (right density-orange line) at 0 (2007-2013)\label{fig:Headcount-Posterior-Density}}

\begin{centering}
\includegraphics[width=15cm,height=10cm]{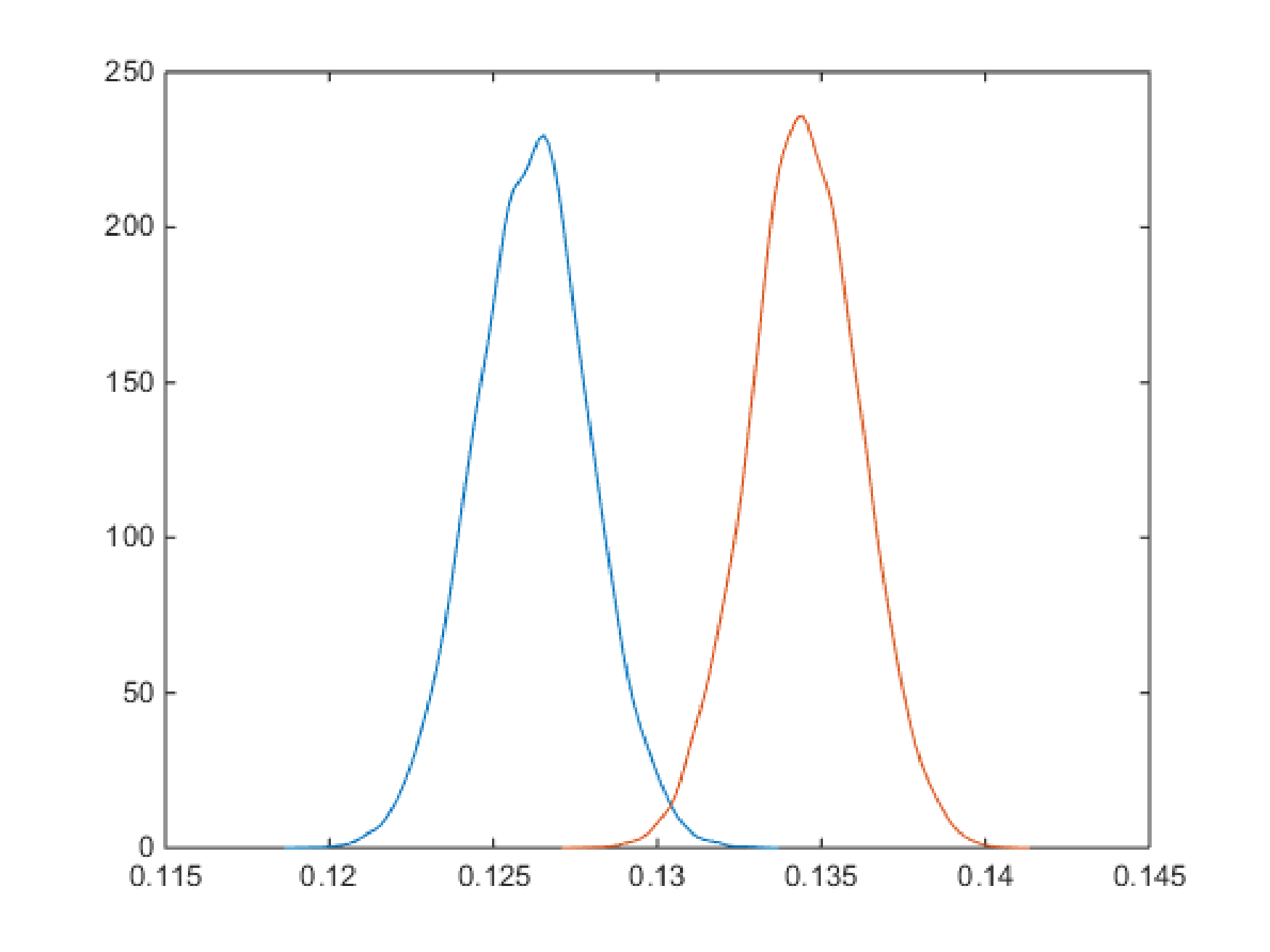}
\par\end{centering}

\end{figure}

Tables \ref{tab:Spearman-Rho-Dependence} and \ref{tab:Shorrocks-(1978a)-Mobility-1}
show the estimate of Spearman's rho dependence and \cite{Shorrocks1978a}
mobility measure. We can see from these two measures that there are
very high values of the mobility measure and very low values of Spearman's
rho dependence measure between 2006-2007 and 2007-2008. This confirms
our previous analysis that in the period 2006-2008  there  is very
high mobility between income earners. Table \ref{tab:Foster's-Chronic-Poverty}
shows the estimates of Foster's chronic poverty measures: duration
adjusted headcount ratio and duration adjusted poverty gap. The two
measures indicate that the chronic poverty is significantly lower
in the  2007-2013 period compared to the 2001-2006 period. The standard of living
in Australia is higher in the period 2007-2013 compared to the period
2001-2006. Furthermore, we can see that the estimates of Spearman's
rho dependence, mobility, and chronic measures are different between
the estimates that take into account the point masses and the estimates
that do not take into account the point masses at zero wages. Figure
\ref{fig:Headcount-Posterior-Density}  shows the posterior densities
of duration adjusted headcount ratio for the years 2007-2013 for the
two estimates. The figure shows that the posterior densities almost do
not overlap, indicating that the two estimates are significantly different.
Therefore, whenever the point masses are present, it is strongly recommended
to incorporate them into the model to guard against biased estimates.

\begin{table}[H]
\caption{Estimates of the Spearman rho dependence measure of the mixture of the Gaussian, Gumbel and Clayton copulas  and 95\% credible intervals (in brackets) \label{tab:Spearman-Rho-Dependence}}

\centering{}%
\begin{tabular}{ccc}
\hline
Period & \multicolumn{1}{c}{Copula (Point Mass)} & \multicolumn{1}{c}{Copula (No Point Mass)}\tabularnewline
\hline
\hline
2001-2002 & $\underset{\left(0.684,0.722\right)}{0.703}$ & $\underset{\left(0.723,0.757\right)}{0.740}
$\tabularnewline
2002-2003 & $\underset{\left(0.700,0.737\right)}{0.719}$ & $\underset{\left(0.726,0.759\right)}{0.743
}$\tabularnewline
2003-2004 & $\underset{\left(0.702,0.739\right)}{0.721}$ & $\underset{\left(0.727,0.759\right)}{0.743}
$\tabularnewline
2004-2005 & $\underset{\left(0.7040,0.741\right)}{0.723}$ & $\underset{\left(0.730,0.763\right)}{0.747}
$\tabularnewline
2005-2006 & $\underset{\left(0.708,0.745\right)}{0.727}$ & $\underset{\left(0.733,0.766\right)}{0.750}
$\tabularnewline
2006-2007 &  $\underset{\left(-0.028,0.068\right)}{0.020}$ & $\underset{\left(-0.020,0.086\right)}{0.030}$\tabularnewline
2007-2008 & $\underset{\left(-0.023,0.073\right)}{0.025}$ & $\underset{\left(-0.013,0.093\right)}{0.037}$\tabularnewline
2008-2009 & $\underset{\left(0.706,0.744\right)}{0.725}$ & $\underset{\left(0.733,0.766\right)}{0.7500}$\tabularnewline
2009-2010 & $\underset{\left(0.716,0.753\right)}{0.735}$ & $\underset{\left(0.741,0.775\right)}{0.758}
$\tabularnewline
2010-2011 & $\underset{\left(0.720,0.758\right)}{0.740}$ & $\underset{\left(0.747,0.781\right)}{0.764}
$\tabularnewline
2011-2012 & $\underset{\left(0.718,0.755\right)}{0.737}$ &  $\underset{\left(0.745,0.778\right)}{0.762}
$\tabularnewline
2012-2013 & $\underset{\left(0.714,0.752\right)}{0.733}$ & $\underset{\left(0.742,0.776\right)}{0.759}
$\tabularnewline
\hline
\end{tabular}
\end{table}

\begin{table}[H]
\caption{Estimates of Shorrocks (1978a) Mobility Measure ($m=5$) of the mixture of the Gaussian, Gumbel and Clayton copulas \label{tab:Shorrocks-(1978a)-Mobility-1}}

\centering{}%
\begin{tabular}{cccc}
\hline
Period & Non-Parametric & \multicolumn{1}{c}{Copula (Point Mass)} & \multicolumn{1}{c}{Copula (No Point Mass)}\tabularnewline
\hline
\hline
2001-2002 & $\underset{\left(0.367,0.466\right)}{0.414}$ & $\underset{\left(0.549,0.588\right)}{0.569}$ & $\underset{\left(0.501,0.534\right)}{0.518}$\tabularnewline
2002-2003 & $\underset{\left(0.361,0.461\right)}{0.411}$ & $\underset{\left(0.508,0.543\right)}{0.526}$ &  $\underset{\left(0.484,0.516\right)}{0.499}$\tabularnewline
2003-2004 & $\underset{\left(0.324,0.409\right)}{0.366}$ & $\underset{\left(0.483,0.516\right)}{0.500}$ & $\underset{\left(0.463,0.495\right)}{0.479}$\tabularnewline
2004-2005 & $\underset{\left(0.341,0.418\right)}{0.380}$ & $\underset{\left(0.473,0.506\right)}{0.489}$ & $\underset{\left(0.450,0.480\right)}{0.465}$\tabularnewline
2005-2006 & $\underset{\left(0.352,0.427\right)}{0.392}$ & $\underset{\left(0.468,0.5000\right)}{0.484}$ & $\underset{\left(0.444,0.475\right)}{0.459}$\tabularnewline
2006-2007 & $\underset{\left(0.974,1.019\right)}{0.996}$ & $\underset{\left(0.957,0.980\right)}{0.969}$ & $\underset{\left(0.878,0.938\right)}{0.918}$\tabularnewline
2007-2008 & $\underset{\left(0.959,1.015\right)}{0.987}$ & $\underset{\left(0.921,0.945\right)}{0.933}$ & $\underset{\left(0.843,0.906\right)}{0.885}$\tabularnewline
2008-2009 & $\underset{\left(0.384,0.441\right)}{0.411}$ & $\underset{\left(0.493,0.526\right)}{0.510}$ & $\underset{\left(0.465,0.495\right)}{0.480}$\tabularnewline
2009-2010 & $\underset{\left(0.350,0.409\right)}{0.380}$ & $\underset{\left(0.482,0.516\right)}{0.500}$ & $\underset{\left(0.449,0.481\right)}{0.465}$\tabularnewline
2010-2011 & $\underset{\left(0.351,0.411\right)}{0.381}$ & $\underset{\left(0.463,0.500\right)}{0.481}$ & $\underset{\left(0.424,0.456\right)}{0.440}$\tabularnewline
2011-2012 & $\underset{\left(0.353,0.405\right)}{0.380}$ & $\underset{\left(0.475,0.510\right)}{0.492}$ &  $\underset{\left(0.437,0.469\right)}{0.453}$\tabularnewline
2012-2013 & $\underset{\left(0.339,0.395\right)}{0.365}$ & $\underset{\left(0.499,0.536\right)}{0.517}$ & $\underset{\left(0.458,0.493\right)}{0.475}$\tabularnewline
\hline
\end{tabular}
\end{table}

\begin{sidewaystable}
\caption{Estimates of Foster's chronic poverty measure of the mixture of the Gaussian, Gumbel and Clayton copulas with 95\% credible intervals (in brackets)\label{tab:Foster's-Chronic-Poverty}}

\centering{}%

\begin{tabular}{ccccc}
\hline
Measure & Period & Non-Parametric & \multicolumn{1}{c}{Copula (Point Mass)} & \multicolumn{1}{c}{Copula (No Point Mass)}\tabularnewline
\hline
\hline
adj. headcount  & 2001-2006 & $\underset{\left(0.193,0.229\right)}{0.211}$ & $\underset{\left(0.187,0.197\right)}{0.192}$ & $\underset{\left(0.197,0.205\right)}{0.201}$\tabularnewline
adj. headcount & 2007-2013 & $\underset{\left(0.120,0.149\right)}{0.135}$ & $\underset{\left(0.123,0.130\right)}{0.126}$ & $\underset{\left(0.131,0.138\right)}{0.135}$\tabularnewline
adj. poverty gap & 2001-2006 & $\underset{\left(0.123,0.150\right)}{0.137}$ & $\underset{\left(0.130,0.137\right)}{0.134}$ & $\underset{\left(0.137,0.144\right)}{0.141}$\tabularnewline
adj. poverty gap & 2007-2013 & $\underset{\left(0.095,0.119\right)}{0.108}$ & $\underset{\left(0.101,0.107\right)}{0.104}$ & $\underset{\left(0.108,0.114\right)}{0.111}$\tabularnewline
\hline

\end{tabular}
\end{sidewaystable}

\section{Conclusion and discussion \label{S: conclusion}}

\textcolor{blue}{The paper shows how to define and derive the density of the observations of a general copula model when
some of the marginals are discrete, some are continuous {\em and} some of the marginals are a mixture of discrete and continuous components.
This is done by carefully defining the likelihood as the
density of the observations with respect to a mixed measure
and  allows us to define the likelihood for general copula models and hence carry out likelihood based inference.
Our work extends in a very general way the current literature on likelihood based inference which focuses
on copulas where each marginal is either discrete or continuous. The inference in the paper is Bayesian and we show how to construct an
efficient MCMC scheme to estimate functionals of the posterior distribution. Although our discussion and examples focus on Gaussian and Archimedean copulas, our treatment is quite general and can be applied as long as it is possible to compute certain marginal and conditional copulas either in closed-form or numerically.}

\textcolor{blue}{Our article can be extended in the following directions.
First, using our definition of the likelihood also enables maximum likelihood type inference using, for example, simulated EM or simulated maximum likelihood.
Second, copulas based on pair-copula constructions \citep[e.g.][]{aas2009pair} or vine copulas \citep[e.g.][]{bedford2002vines} lend themselves well to our approach because
the methods in this paper apply to arbitrary copulas with  the only requirement that it is possible
to write down several conditional marginal copulas and copula densities and being able to compute those either analytically or numerically.
Third, by using pseudo marginal methods \citep[e.g.][]{Andrieu:2010}, our methodology can also be extended to the case where the case where the likelihood of the copula can only be estimated unbiasedly, rather than evaluated. We leave all such extensions to future work.}

Our article illustrates the methodology and algorithms by applying them to
estimate a multivariate income dynamics model.
Examples of further possible applications arise from any setup where one or more  of the
following variables are present: wages (where there are points of probability mass at the
minimum wage) individual sales figures, where there is a \textcolor{blue}{point of probability mass} at 0 (many
individuals deciding not to purchase) and a smooth distribution above that
point (corresponding to a continuum of price figures). Another
interesting potential application is to extend the general truncated/censored
variable models in econometrics to a copula framework, e.g., for multivariate tobit and
sample selection models.

\section*{Acknowledgement}
We would like to thank two anonymous referees and the associate editor for suggestions that helped improve the clarity of the paper.
The research of David Gunawan and Robert Kohn was partially supported by an Australian Research Council Discovery Grant
DP150104630.

\bibliographystyle{asa}
\bibliography{references_v1}

\begin{appendices}

\section{Difference operator notation\label{app_diference_operator}}

Since the difference operator notation can be easily confusing, it is
useful to adopt the convention below. The notation has two
 components:
\begin{enumerate}
  \item Whenever the $\bigtriangleup$ operators are applied to a function, an
  indexing is used to make the domain of the function clear.

  \item A dot  marks  the position of the variables that are being
  differenced.
\end{enumerate}
Here are some examples to illustrate the use of that notation.
\begin{itemize}
  \item Consider a function $g (x)$ where $x$ is a scalar. Then
  $\bigtriangleup_a^b g_x (\cdot)$ defines
  \[ \bigtriangleup_a^b g_x (\cdot) := g (b) - g (a) \]
  \item Consider a function $g (x, y)$ where both $x$ and $y$ are scalars. By
  $\bigtriangleup_a^b g_{x, y} (\cdot, z)$ we mean that the differencing is only
  applied to $x$ while the second argument is fixed at $y = z$, that is
  \[ \bigtriangleup_a^b g_{x, y} (\cdot, z) := g (b, z) - g (a, z) \]
  \item Consider a function $g (\ensuremath{\boldsymbol{x}})$ where
  $\ensuremath{\boldsymbol{x}}$ is two-dimensional. By
  $\bigtriangleup_{\ensuremath{\boldsymbol{a}}}^{\ensuremath{\boldsymbol{b}}}
  g_{\ensuremath{\boldsymbol{x}}} (\cdot)$, we mean
  \begin{eqnarray*}
    \bigtriangleup_{\ensuremath{\boldsymbol{a}}}^{\ensuremath{\boldsymbol{b}}}
    g_{\ensuremath{\boldsymbol{x}}} (\cdot) & = & \bigtriangleup_{a_1}^{b_1}
    \bigtriangleup_{a_2}^{b_2} g_{\ensuremath{\boldsymbol{x}}} (\cdot)\\
    & = & \bigtriangleup_{a_1}^{b_1} (g_{x_1, x_2} (\cdot, b_2) - g_{x_1, x_2}
    (\cdot, a_2))\\
    & = & g (b_1, b_2) - g (b_1, a_2) - g (a_1, b_{2 }) + g (a_1, a_2)
  \end{eqnarray*}
  \item Consider a function $g (\ensuremath{\boldsymbol{x}},
  \ensuremath{\boldsymbol{y}})$. If the differencing is applied to
  $\ensuremath{\boldsymbol{y}}$ and not $\ensuremath{\boldsymbol{x}}$, and if
  $\ensuremath{\boldsymbol{y}}$ is two-dimensional, then
  $\bigtriangleup_{\ensuremath{\boldsymbol{a}}}^{\ensuremath{\boldsymbol{b}}}
  g_{\ensuremath{\boldsymbol{x}}, \ensuremath{\boldsymbol{y}}}
  (\ensuremath{\boldsymbol{x}}, \cdot)$ means
  \begin{eqnarray*}
    \bigtriangleup_{\ensuremath{\boldsymbol{a}}}^{\ensuremath{\boldsymbol{b}}}
    g_{\ensuremath{\boldsymbol{x}}, \ensuremath{\boldsymbol{y}}}
    (\ensuremath{\boldsymbol{x}}, \cdot) & := & \bigtriangleup_{a_1}^{b_1}
    \bigtriangleup_{a_2}^{b_2} g_{\ensuremath{\boldsymbol{x}},
    \ensuremath{\boldsymbol{y}}} (\ensuremath{\boldsymbol{x}}, \cdot)\\
    & = & \bigtriangleup_{a_1}^{b_1} (g_{\ensuremath{\boldsymbol{x}}, y_1,
    y_2} (\ensuremath{\boldsymbol{x}}, \cdot, b_2) -
    g_{\ensuremath{\boldsymbol{x}}, y_1, y_2} (\ensuremath{\boldsymbol{x}}, \cdot,
    b_2))\\
    & = & g (\ensuremath{\boldsymbol{x}}, b_1, b_2) - g
    (\ensuremath{\boldsymbol{x}}, a_1, b_2) - g (\ensuremath{\boldsymbol{x}},
    a_2, b_1) + g (\ensuremath{\boldsymbol{x}}, a_1, a_2)
  \end{eqnarray*}
\end{itemize}

\section{Deriving the likelihood and the conditional density\label{app_integration}}

This appendix deals with densities defined with respect to mixed measures. Such densities are formally defined by Radon-Nikodym derivatives. In particular, we
obtain the joint density~\eqref{JointDensity} of $\bsX$ and $\bsU$ and the
corresponding mixed measure. We then show how to obtain the closed-form formulas for the densities
\eqref{likelihood} and \eqref{LatentConditionalDistribution}, and their corresponding mixed measures, from the density~\eqref{JointDensity}.

We need the following three elementary lemmas to obtain the results. They are likely to be
 known in the literature, but we include their proofs for completeness.
\begin{lemma}  \label{L: deriv F}
  Let $F (\ensuremath{\boldsymbol{x}}, \ensuremath{\boldsymbol{y}})$ be the
  distribution function of an absolutely continuous random vector
  $(\ensuremath{\boldsymbol{X}}', \ensuremath{\boldsymbol{Y}}')'$ where
  $\ensuremath{\boldsymbol{x}} \in \mathbbm{R}^k$ and
  $\ensuremath{\boldsymbol{y}} \in \mathbbm{R}^p$.  Then,
  \[ \frac{\partial^k F (\ensuremath{\boldsymbol{x}},
     \ensuremath{\boldsymbol{y}})}{\partial x_1 \cdots \partial x_k} = F
     (\ensuremath{\boldsymbol{y}}|\ensuremath{\boldsymbol{x}}) f
     (\ensuremath{\boldsymbol{x}}),  \]
  where $F (\ensuremath{\boldsymbol{y}}|\ensuremath{\boldsymbol{x}})$ and $f
  (\ensuremath{\boldsymbol{x}})$ are respectively the distribution
  function of $\ensuremath{\boldsymbol{Y}}$ conditional on
  $\ensuremath{\boldsymbol{X}}=\ensuremath{\boldsymbol{x}}$ and the density of
  $\ensuremath{\boldsymbol{X}}$.
  Similarly, in an obvious notation,
  \[ \frac{\partial^p F (\ensuremath{\boldsymbol{x}},
     \ensuremath{\boldsymbol{y}})}{\partial y_1 \cdots \partial y_p} = F
     (\ensuremath{\boldsymbol{x}}|\ensuremath{\boldsymbol{y}}) f
     (\ensuremath{\boldsymbol{y}}). \]

\begin{proof}
  The identity comes from
  \[ \frac{\partial^p}{\partial y_1 \cdots \partial y_p} F
     (\ensuremath{\boldsymbol{y}}|\ensuremath{\boldsymbol{x}}) = f
     (\ensuremath{\boldsymbol{y}}|\ensuremath{\boldsymbol{x}}) = \frac{f
     (\ensuremath{\boldsymbol{y}}, \ensuremath{\boldsymbol{x}})}{f
     (\ensuremath{\boldsymbol{x}})} = \frac{\frac{\partial^{p + k} F
     (\ensuremath{\boldsymbol{x}}, \ensuremath{\boldsymbol{y}})^{}}{\partial
     y_1 \cdots \partial y_p \partial x_1 \cdots \partial x_k}}{f
     (\ensuremath{\boldsymbol{x}})} = \frac{\partial^p}{\partial y_1 \cdots
     \partial y_p} \left( \frac{\frac{\partial^k F
     (\ensuremath{\boldsymbol{x}}, \ensuremath{\boldsymbol{y}})^{}}{\partial
     x_1 \cdots \partial x_k}}{f (\ensuremath{\boldsymbol{x}})} \right) . \]

\end{proof}
\end{lemma}

The next lemma is an immediate consequence of the previous lemma.
\begin{lemma}\label{L: lemma F 2 }
  Let $f (\ensuremath{\boldsymbol{x}}, \ensuremath{\boldsymbol{y}})$ be the
  density of an absolutely continuous random vector
  $(\ensuremath{\boldsymbol{X}}', \ensuremath{\boldsymbol{Y}}')'$ where
  $\ensuremath{\boldsymbol{x}} \in \mathbbm{R}^k$ and
  $\ensuremath{\boldsymbol{y}} \in \mathbbm{R}^p$ then
  \[ \int_{a_1}^{b_1} \cdots \int_{a_k}^{b_k} f (\ensuremath{\boldsymbol{x}},
     \ensuremath{\boldsymbol{y}}) \mathrm{d} x_1 \ldots \mathrm{d} x_k =
     \bigtriangleup_{a_1}^{b_1} \cdots \bigtriangleup_{a_k}^{b_k}
     F_{\ensuremath{\boldsymbol{Y}}|\ensuremath{\boldsymbol{X}}}
     (\ensuremath{\boldsymbol{y}}|.) f_{\ensuremath{\boldsymbol{X}}} (.) \]
  where $F (\ensuremath{\boldsymbol{y}}|\ensuremath{\boldsymbol{x}})$ and $f
  (\ensuremath{\boldsymbol{x}})$ are respectively the conditional distribution
  function of $\ensuremath{\boldsymbol{Y}}$ on
  $\ensuremath{\boldsymbol{X}}=\ensuremath{\boldsymbol{x}}$ and the density of
  $\ensuremath{\boldsymbol{X}}$.

\begin{proof}
  Write the density function
  \begin{eqnarray*}
    f (\ensuremath{\boldsymbol{x}}, \ensuremath{\boldsymbol{y}}) & = &
    \frac{\partial^{p + k} F (\ensuremath{\boldsymbol{x}},
    \ensuremath{\boldsymbol{y}})^{}}{\partial y_1 \cdots \partial y_p \partial
    x_1 \cdots \partial x_k}\\
    & = & \frac{\partial^k}{\partial x_1 \cdots \partial x_k} \left(
    \frac{\partial^p F (\ensuremath{\boldsymbol{x}},
    \ensuremath{\boldsymbol{y}})^{}}{\partial y_1 \cdots \partial y_p}
    \right)\\
    & = & \frac{\partial^k}{\partial x_1 \cdots \partial x_k} (F
    (\ensuremath{\boldsymbol{x}}|\ensuremath{\boldsymbol{y}}) f
    (\ensuremath{\boldsymbol{y}}))
  \end{eqnarray*}
  where the last line follows from the previous lemma.
  The desired result follows by an application of the fundamental theorem of
  calculus.
\end{proof}
\end{lemma}

\begin{lemma} \label{L: elementary}
Suppose that $U$ is uniformly distributed on $[0,1]$.
\begin{enumerate}
\item [(i)]
Suppose that $X$ is a univariate random variable with CDF $F(x)$ that has an inverse and a density $f(x)$.
Then, $\mrd u \delta_{F^{-1}(u)} (\mrd x) = \delta_{F(x)}(\mrd u) f(x) \mrd x $, where $\mrd u, \mrd x$ are Lebesgue measures.
\item [(ii)] Suppose that $X$ is a discrete univariate random variable with support on the discrete set $I=\{x\}$. Then,
$\mrd u \delta_{\{F(x^-) \leq u < F(x)\}  }(\mrd x) = {\cal{I}} \{u: F(x^-) \leq u < F(x)\} \mrd u \delta_I(\mrd x)$
\end{enumerate}
The proofs of parts~(i) and (ii) are in Section~\ref{proof: proof of lemma 3}.
\end{lemma}

Suppose that the indices $\M_\C$  correspond to the continuous random variables, the indices $\M_\D $ to the
discrete random variables and the indices $\M_\J $ to a mixture of discrete and continuous random variables. We define the joint density of
$\bsX$ and $\bsU$ as
\begin{align}\label{eq: full density x and u}
f(\bx,\bu) & := c(\bu) \prod_{j \in \M_\C } {\cal{I}}(u_j = F_j(x_j))\prod_{j \in \M_\D }{\cal{I}}( F_j(x_j^-) \leq u_j < F_j(x_j) )\times  \nonumber\\
&  \prod_{j \in \M_\J}
\big ( {\cal{I}}(u_j = F_j(x_j))  +  {\cal{I}}( F_j(x_j^-) \leq u_j < F_j(x_j) ) \big)
\end{align}
with respect to the measure
\begin{align}\label{eq: full measure x and u}
& \mrd \bu  \prod_{j \in \M_\C } \delta_{F_j^{-1}(u_j)}(\mrd x_j) \prod_{j \in \M_\D} \delta_{F_j(x_j^-) \leq u_j < F_j(x_j)}(\mrd x_j) \times \nonumber \\
 & \prod_{j \in \M_\J} \big ( {\cal I }(u_j = F_j(x_j)) \mrd x_j
 + {\cal{I}}( F_j(x_j^-) \leq u_j < F_j(x_j) )\delta_{F_j(x_j^-) \leq u_j < F_j(x_j)}(\mrd x_j)
 \big )
 \end{align}
\begin{lemma} \label{L: mixed density in x and u}
\begin{enumerate}
\item [(i)]
Equation \eqref{JointDensity} gives the joint density  of $\bsX$ and $\bsU$ at a given value $\bsX = \bsx$ and $\bsU = \bsu$.
 \item [(ii)]
Equation~\eqref{likelihood} is the marginal density of $\bsX$ at $\bsX = \bsx$.
 \item [(iii)]  Equation~\eqref{LatentConditionalDistribution} is the conditional density of $\bsU_{\D(\bsx)}$ given $\bsX = \bsx$.
 \end{enumerate}
 \begin{proof}
Part~(i) follows directly from \eqref{eq: full density x and u} and \eqref{eq: full measure x and u}. Part~(ii) follows by integrating out $\bsu$ using
Lemma~\ref{L: lemma F 2 }.
Part~(iii) follows from Lemma~ \ref{L: elementary}.
\end{proof}
\end{lemma}

\end{appendices}

\clearpage
\renewcommand{\theequation}{S\arabic{equation}}
\renewcommand{\thesection}{S\arabic{section}}
\renewcommand{\theproposition}{S\arabic{proposition}}
\renewcommand{\theassumption}{S\arabic{assumption}}
\renewcommand{\thelemma}{S\arabic{lemma}}
\renewcommand{\thecorollary}{S\arabic{corollary}}
\renewcommand{\thealgorithm}{S\arabic{algorithm}}
\renewcommand{\thefigure}{S\arabic{figure}}
\renewcommand{\thetable}{S\arabic{table}}
\renewcommand{\thepage}{S\arabic{page}}
\renewcommand{\thetable}{S\arabic{table}}
\renewcommand{\thepage}{S\arabic{page}}
\setcounter{page}{1}
\setcounter{section}{0}
\setcounter{equation}{0}
\setcounter{algorithm}{0}
\setcounter{table}{0}
\title{\Huge  \sf Supplement to \lq Mixed marginal Coupula Modeling\rq}

\renewcommand\Authands{ and }

\maketitle

\section{Density, Conditional Distribution Function, and MCMC Sampling Methods
for the  Gaussian, Gumbel, and Clayton Copulas \label{app_sampling_schemes}}

\subsection{Gaussian copula}
The Gaussian copula distribution and density function are given by
\cite{Song2000} as
$
C\left(u_{1},u_{2},...,u_{m};\Gamma\right)=\Phi_{m}\left(y_{1}^{*},y_{2}^{*},...,y_{m}^{*};\Gamma\right)
$
and
\begin{align}
c\left(u_{1},u_{2},...,u_{m};\Gamma\right) & = 
   |\Gamma|^{-1/2}\exp\left\{ -\frac{1}{2}\mathbf{y}^{*'}\left(\Gamma^{-1}-I\right)\mathbf{y}^{*}\right\},  \label{eq:gaussian copula density}
\end{align}
where $\mathbf{y}^{*}=\left(y_{1}^{*},y_{2}^{*},...,y_{m}^{*}\right)^{'}$
and $y_{j}^{*}=\Phi^{-1}\left(F_{j}\left(y_{j};\boldsymbol{\theta}_{j}\right)\right)$
is the transformed Gaussian copula data; $\Phi_{m}\left(\right)$
is the distribution function of the standard $m$- dimensional multivariate
Gaussian distribution $N\left(\mathbf{0},\Gamma\right)$ and $\Gamma$
is a correlation matrix. The correlation matrix $\Gamma$ captures
the dependence among random variables $\mathbf{y}^{*}=\left(y_{1}^{*},y_{2}^{*},...,y_{m}^{*}\right)^{'}$.
There are $m\left(m-1\right)/2$ unknown parameters in the correlation
matrix $\Gamma$. We can generate a random sample from the
Gaussian copula as follows,
\begin{itemize}
\item Generate $z_{1},...,z_{m}$ from $N\left(0,\Gamma\right)$
\item Compute a vector $\mathbf{u}=\left(\Phi\left(z_{1}\right),...,\Phi\left(z_{m}\right)\right)^{'}$
\item Compute $\mathbf{x}=\left(F_{1}^{-1}\left(u_{1}\right),...,F_{T}^{-1}\left(u_{m}\right)\right)^{'}$
\end{itemize}

\subsection{Clayton and Gumbel copulas}\label{SS: Clayton and Gumbel copulas}
The material in this section is covered in more detail in \cite{Hofert2012} and \cite{Cherubini2004}.
We consider a strict generator function
\[
\psi\left(u\right):\left[0,1\right]\rightarrow\left[0,\infty\right]
\]
which is continuous and strictly decreasing, with $\psi^{-1}$ completely monotonic
on $\left[0,\infty\right]$. Then, the class of Archimedean copulas
consists of copulas of the form \citep{Cherubini2004}
\[
C\left(\mathbf{u}\right)=C\left(u_{1},...,u_{m}\right)=\psi^{-1}\left(\psi\left(u_{1}\right)+...+\psi\left(u_{m}\right)\right).
\]
A function $\psi$ on $\left[0,\infty\right]$ is the Laplace transform
of a CDF $F$ if and only if $\psi$ is a completely monotonic and
$\psi\left(0\right)=1$ and $\psi\left(\infty\right)=0$. Applying
Bayesian methodology requires an efficient strategy to evaluate the
density or the log-density of the parametric Archimedean copula family
to be estimated. Although the density of an Archimedean copula has an
explicit form in theory,  it is often difficult to compute since computing
the required derivatives is known to be extremely challenging, especially
in high dimensional applications. \cite{Hofert2012} gives explicit
formulae for the generator derivatives of the Archimedean family
in any dimension. They also give an explicit formula for the density
of some well-known Archimedean copulas, such as Ali-Mikhail-Haq, Clayton,
Frank, Gumbel, and Joe copulas.

The generator for the Clayton copula is $\psi\left(u\right)=u^{-\theta}-1$
with $\psi^{-1}\left(t\right)=\left(1+t\right)^{-\frac{1}{\theta}}$.
The CDF of the Clayton $m$-copula is
\[
C\left(\mathbf{u}\right)=\left[\sum_{i=1}^{m}u_{i}^{-\theta}-m+1\right]^{-\frac{1}{\theta}},\theta>0 .
\]
The dependence parameter $\theta$ is defined on the interval $\left(0,\infty\right)$.
The Clayton copula favors data which exhibits strong lower tail dependence
and weak upper tail dependence and thus is an appropriate
choice of model if the data exhibits strong correlation at lower values
and weak correlation at higher values. The density of the Clayton $m$-copula
is
\[
c\left(\mathbf{u}\right)=\prod_{k=0}^{m-1}\left(\theta k+1\right)\left(\prod_{j=1}^{m}u_{j}\right)^{-\left(1+\theta\right)}\left(\sum_{i=1}^{m}u_{i}^{-\theta}-m+1\right)^{-\left(m+\frac{1}{\theta}\right)}.
\]
The generator of the Gumbel copula is $\psi\left(u\right)=\left(-\log\left(u\right)\right)^{\theta}$
with $\psi^{-1}\left(t\right)=\exp\left(-t^{\frac{1}{\theta}}\right)$.
The Gumbel $m$-copula CDF is
\[
C\left(\mathbf{u}\right)=\exp\left\{ -\left[\sum_{i=1}^{m}\left(-\log\left(u_{i}\right)\right)^{\theta}\right]^{\frac{1}{\theta}}\right\}.
\]
The dependence parameter $\theta$ is defined on the $\left[1,\infty\right)$
interval, where a value 1 represents the independence case. The Gumbel copula
is an appropriate choice of model if the data exhibit weak correlation
at lower values and strong correlation at the higher values. The density
of the Gumbel $m$-copula is
\begin{align*}
  c \left( \mathbf{u} \right)  =  \theta^m \exp \left\{ - \left[ \sum_{i =
  1}^m (- \log (u_i))^{\theta} \right]^{\frac{1}{\theta}} \right\}
   \times &  \frac{\prod_{j = 1}^m (- \log u_j)^{\theta - 1}}{\left( \sum_{j
  = 1}^m (- \log (u_j))^{\theta} \right)^m  \prod_{j = 1}^m u_j} \\
  & \times  P_{d, \theta}^G \left( \left[ \sum_{j = 1}^m (- \log
  (u_j))^{\theta} \right]^{\frac{1}{\theta}} \right)
\end{align*}
where,
\[
P_{m,\theta}^{G}\left(x\right)=\sum_{k=1}^{m}a_{mk}^{G}\left(\theta\right)x^{k},
\]
and
\[
a_{mk}^{G}\left(\theta\right)=\frac{m!}{k!}\sum_{j=1}^{k}\left(\begin{array}{c}
k\\
j
\end{array}\right)\left(\begin{array}{c}
j/\theta\\
m
\end{array}\right)\left(-1\right)^{m-j}.
\]
 \cite{Marshall1988} proposed the following algorithm for sampling a $m$-dimensional exchangeable
Archimedean copula with generator $\psi$.
\begin{itemize}
\item Sample $V\sim F=LS^{-1}\left(\psi^{-1}\right)$, where $LS^{-1}$
denotes the inverse Laplace-Stieljes transform of $\psi^{-1}$.

\begin{itemize}
\item For the Clayton copula, $F=\varGamma\left(1/\theta,1\right)$, where $\varGamma\left(c,d\right)$
denotes the Gamma distribution with shape parameter $c\in\left(0,\infty\right)$,
scale parameter $d\in\left(0,\infty\right)$
\item \textcolor{blue}{For the Gumbel copula, $F={\rm Stable}\left(1/\theta,1,\left(\cos\left(\frac{\pi}{2\theta}\right)\right)^{\theta},1;1\right)$,
\newline where $\rm{Stable}\left(\alpha_{st},\beta_{st},\gamma_{st},\delta_{st};1\right)$
denotes the Stable distribution with exponent $\alpha_{st}\in\left(0,2\right]$,
skewness parameter $\beta_{st}\in\left[-1,1\right]$, scale parameter
$\gamma_{st}\in\left[0,\infty\right)$, and location parameter $\delta_{st}\in \mathbb{R}$
\citep{Nolan:2007}.}

\end{itemize}
\item Sample iid $X_{j}\sim U\left[0,1\right]$ for $j=1,...,m$
\item Set $U_{j}=\psi\left(\frac{-\log\left(X_{j}\right)}{V}\right)$, for
$t=1,...,T$
\end{itemize}

\subsection{Conditional posterior of the Gaussian Copula Parameters}

At each iteration of the MCMC sampling scheme, the correlation matrix $\Gamma$ of the Gaussian copula is generated
conditional on the transformed Gaussian copula variables $\mathbf{y}^{*}=\left\{ y_{ij}^{*};i=1,...,n;\;j=1,...,m\right\} $.
\cite{Danaher2011} proposed the following representation of $\Gamma$,
\[
\Gamma:={\rm diag}\left(\Sigma\right)^{-1/2}\;\Sigma\;\rm{diag}\left(\Sigma\right)^{-1/2},
\]
where $\Sigma$ is a non-unique positive definite matrix and ${\rm diag}(\Sigma)$
is a diagonal matrix comprising the leading diagonal of $\Sigma$. The
matrix $\Sigma$ is further decomposed into $\Sigma=R^{'}R$, with
$R$  an upper triangular Cholesky factor. If we set the leading
diagonal of $R$ to ones, this leaves $m\left(m-1\right)/2$ unknown
elements of $R$, matching the number of unknown elements of $\Gamma$,
thus identifying the representation. The upper triangular elements
of $R$ are unconstrained. The transformation described above ensures
that the correlation matrix $\Gamma$ remains a positive definite
matrix, regardless of the values of $R$.

The following steps generate each element of $R$:
\begin{enumerate}
\item Generate the $r_{j^{*}j}$ element of the matrix $R$ using a random-walk
Metropolis step for $j^{*}=1,...,m$ and $j=2,...,m$ with
$j^{*}<j$.
\item Compute $\Sigma=R^{'}R$
\item Compute the correlation matrix $\Gamma= {\rm diag} \left(\Sigma\right)^{-1/2}\Sigma {\rm diag} \left(\Sigma\right)^{-1/2}$
\end{enumerate}
To explain step 1 in more detail, the conditional posterior $r_{j^{*}j}|\left\{ R\setminus r_{j^{*}j}\right\} ,\mathrm{\mathbf{y}}^{*},\mathbf{y}$
is given by
\begin{eqnarray*}
p\left(r_{j^{*}j}|\left\{ R\setminus r_{j^{*}j}\right\} ,\mathrm{\mathbf{y}}^{*},\mathbf{y}\right) & \propto & p\left(\mathrm{\mathbf{y}}|\mathbf{y}^{*}\right)p\left(\mathbf{y}^{*}|R\right)p\left(r_{j^{*}j}\right)\\
 & \propto & \prod_{i=1}^{n}|\Gamma|^{-\frac{n}{2}}\exp\left\{ -\frac{1}{2}\mathbf{y}_{i}^{*'}\left(\Gamma^{-1}-I\right)\mathbf{y}_{i}^{*}\right\} p\left(r_{j^{*}j}\right),
\end{eqnarray*}
with $p\left(r_{j^{*}j}\right)\propto1$ for all elements of $R$.
Here, $\left\{ A\setminus B\right\} $ means $A$ with the parameters
$B$ omitted. First, we generate a new proposal value, $r_{j^{*}j}^{*}$,
from a candidate density $N\left(r_{j^{*}j},\sigma\right)$, where
$r_{j^{*}j}$ is the previous iterate value and $\sigma$ is the pre-specified
standard deviation of a normal distribution specified to obtain a reasonable acceptance
rate of 0.3-0.4. The new value $r_{j^{*}j}^{*}$ is accepted with probability

\[
\alpha=\min\left\{ 1,\frac{p\left(r_{j^{*}j}^{*}|\left\{ R\setminus r_{j^{*}j}\right\} ,\mathrm{\mathbf{y}}^{*},\mathbf{y}\right)}{p\left(r_{j^{*}j}|\left\{ R\setminus r_{j^{*}j}\right\} ,\mathrm{\mathbf{y}}^{*},\mathbf{y}\right)}\right\}.
\]
We draw a random variable $u$ from $U\left(0,1\right)$; if $u<\alpha$, then
the new value of $r_{j^{*}j}$ is accepted, otherwise the old value
of $r_{j^{*}j}$ is retained. This algorithm is used to generate all
of the upper triangular elements of $R$, one at a time.

\subsection{Generation  of the latent variables} \label{SS: latent marginals}
The following algorithm can be used to generate the latent variables
one margin at a time.
\begin{itemize}
\item In the income application, the point mass occurs at zero wages.
\item For $j=1,...,m$

\begin{itemize}
\item for $i=1,...,n$

\begin{itemize}
\item if $y_{ij}=0$
\item Compute $A_{ij}=C_{j|\left\{ 1,...,m\right\} \setminus j a}\left(b_{i,j}|\left\{ u_{i1},...,u_{ij}\right\} \setminus u_{ij},\phi\right)$,
then generate $w_{ij}\sim Uniform\left(0,A_{ij}\right)$
\item Compute $u_{ij}=C_{j|\left\{ 1,...,m\right\} \setminus j}^{-1}\left(w_{ij}|\left\{ u_{i1},...,u_{im}\right\} \setminus u_{ij},\phi\right)$
\end{itemize}
\end{itemize}
\end{itemize}

\section{A trivariate example\label{S: trivariate example}}

This appendix uses a three dimensional example to illustrate the methods as  some of the more complicated aspects of the methods
may not be apparent in the two dimensional Example~\ref{ex: example 1} discussed in Section~\ref{S: likelihood definition}.
For brevity, the derivation is less detailed than that for Example~\ref{ex: example 1}.

Let $X_1$ have a distribution that is a mixture of two points of probability
mass at zero and one and a normal distribution, that is let $X_1$ has the
distribution function
\[ F_1 (x_1) = (1 - \pi_1 - \pi_2) \Phi (x_1) + \pi_1 \mathcal{I} (x_1
   \geqslant 0) + \pi_2 \mathcal{I} (x_1 \geqslant 1), \]
where $\Phi$ is the distribution function of a standard normal random
variable.
Let $X_2$ be a standard normal with a point of probability mass at $0$ and
with distribution function
\[ F_2 (x_2) = (1 - \pi) \Phi (x) + \pi \mathcal{I} (x_2 \geqslant 0) . \]
Finally, let $X_3$ be a binary random variable.

This results in the following
\[ \mathcal{C} (x_1, x_2, x_3) = \left\{ \begin{array}{lll}
     \{ 1, 2 \} & \mathrm{if} & x_1 \not\in \{ 0, 1 \} \quad \mathrm{ and }\quad  x_2 \not\in \{
     0 \}\\
     \{ 1 \} &  & x_1 \not\in \{ 0, 1 \} \quad \mathrm{ and }\quad  x_2 \in \{ 0 \}\\
     \{ 2 \} &  & x_1 \in \{ 0, 1 \} \quad \mathrm{ and }\quad  x_2 \not\in \{ 0 \}\\
     \varnothing &  & x_1 \in \{ 0, 1 \} \quad \mathrm{ and } \quad x_2 \in \{ 0 \}
   \end{array} \right. \]
and $\mathcal{D} (x_1, x_2, x_3) = \{ 1, 2, 3 \} \backslash \mathcal{C} (x_1,
x_2, x_3)$. Notice that $\{ 3 \} \subset \mathcal{D} (x_1, x_2, x_3)$ always
holds in this example.

The marginal density of $\mathbf{X}$ (Eq.~\ref{likelihood}  in the paper) is
\begin{enumerate}
  \item Case 1: $\mathcal{C} (x_1, x_2, x_3) = \{ 1, 2 \}$.
  \begin{eqnarray*}
    f (\mathbf{x}) & = & c_{\{ 1, 2 \}} (\mathbf{b}_{\{ 1, 2 \}}) f_1
    (x_1) f_2 (x_2) \bigtriangleup_{a_3}^{b_3} C_{3|1, 2} (\cdot|\mathbf{b}_{\{
    1, 2 \}})\\
    & = & c_{\{ 1, 2 \}} (\mathbf{b}_{\{ 1, 2 \}}) f_1 (x_1) f_2 (x_2)
    (C_{3|1, 2} (b_3 |\mathbf{b}_{\{ 1, 2 \}}) - C_{3|1, 2} (a_3
    |\mathbf{b}_{\{ 1, 2 \}}))
  \end{eqnarray*}
  \item Case 2: $\mathcal{C} (x_1, x_2, x_3) = \{ 1 \}$
  \begin{eqnarray*}
    f (\mathbf{x}) & = & c_1 (b_1) f_1 (x_1)
    \bigtriangleup^{\mathbf{b}_{\{ 2, 3 \}}}_{\mathbf{a}_{\{ 2, 3 \}}}
    C_{2, 3|1} (\cdot|b_1)\\
    & = & f_1 (x_1) \times\\
    &  & [C_{2, 3|1} (b_2, b_3 |b_1) - C_{2, 3|1} (b_2, a_3 |b_1) - C_{2,
    3|1} (a_2, b_3 |b_1) + C_{2, 3|1} (a_2, a_3 |b_1)]
  \end{eqnarray*}
  where the second line follows from $c_1 (b_1) = 1$ (as all one-dimensional
  marginals are uniform).

  \item Case 3: $\mathcal{C} (x_1, x_2, x_3) = \{ 2 \}$
  \begin{eqnarray*}
    f (\mathbf{x}) & = & c_2 (b_2) f_2 (x_2)
    \bigtriangleup^{\mathbf{b}_{\{ 1, 3 \}}}_{\mathbf{a}_{\{ 1, 3 \}}}
    C_{1, 3|2} (\cdot|b_2)\\
    & = & f_2 (x_2) \times\\
    &  & [C_{1, 3|2} (b_1, b_3 |b_2) - C_{1, 3|2} (b_1, a_3 |b_2) - C_{1,
    3|2} (a_1, b_3 |b_2) + C_{1, 3|2} (a_1, a_3 |b_2)]
  \end{eqnarray*}
  \item Case 4: $\mathcal{C} (x_1, x_2, x_3) = \varnothing$.
  \begin{align*}
    f (\mathbf{x}) & = C (b_1, b_2, b_3) - C (a_1, b_2, b_3)
    - C (b_1, a_2, b_3) - C (b_1, b_2, a_3)\\
    &   + C (a_1, a_2, b_3) + C (a_1, b_2, a_3)
     + C (b_1, a_2, a_3) - C (a_1, a_2, a_3)
  \end{align*}
\end{enumerate}
In all the above, $b_j = F_j (x_j)$ and $a_j = F_j (x_j^-)$.

\section{Proof of Lemma 3\label{proof: proof of lemma 3}}

\begin{proof}
\begin{itemize}
\item[ (i)]Suppose that $X$ is a univariate absolutely continuous random variable. Then
the cumulative distribution function of $X$ is a strictly increasing $F$ and
$U = F (X)$ will be uniformly distributed on the unit interval. The measure
induced by $(X, U)$ is denoted by $\delta_{F (x)} (\mathd u) f (x) \mathd x$

Let $h(x,u)$ be an integrable function of $x$ and $u$. Then, it is straightforward to check that
\begin{align*}
\int \int h(x,u) f(x)\delta_{F (x)} (\mathd u) \mathd x & = \int h(x, F(x) ) \mathd x = \int \int h(x,u)
 \mathd u \delta_{F^{-1} (u)} (\mathd x)
 \end{align*}

\item [ (ii)] The proof follows from the basic properties of a double integral because we can swap the order of integration.
More formally, the proof follows from Theorem 3.1 (4) p.111 of \cite{shorack2000probability}.
\end{itemize}
\end{proof}
\section{Some extra empirical results} \label{app: extra empirical results}
This appendix includes additional plots for the analysis of the income dynamics data.
\begin{table}[H]
\caption{Estimates of transition probabilities estimates of the mixture of the Gaussian, Gumbel and Clayton copulas taking into account the point masses at zero incomes and  95\% credible intervals (in brackets) \label{tab:Transition-Probability aug-2}}

\centering{}%
\begin{tabular}{ccccc}
\hline
Transition & \multicolumn{2}{c}{0 to Positive Wages} & \multicolumn{2}{c}{Positive to 0 Wages}\tabularnewline
\hline
\hline
 & Non-Parametric & Copula & Non-Parametric & Copula\tabularnewline
\hline
2001-2002 & $\underset{\left(0.388,0.531\right)}{0.465}$ & $\underset{\left(0.477,0.507\right)}{0.492}$ & $\underset{\left(0.031,0.050\right)}{0.040}$ & $\underset{\left(0.042,0.046\right)}{0.044}$\tabularnewline
2002-2003 & $\underset{\left(0.290,0.434\right)}{0.361}$ & $\underset{\left(0.391,0.423\right)}{0.407}$ & $\underset{\left(0.035,0.055\right)}{0.045}$ & $\underset{\left(0.048,0.052\right)}{0.050}$\tabularnewline
2003-2004 & $\underset{\left(0.234,0.379\right)}{0.313}$ & $\underset{\left(0.383,0.414\right)}{0.398}$ & $\underset{\left(0.032,0.053\right)}{0.043}$ & $\underset{\left(0.051,0.054\right)}{0.053}$\tabularnewline
2004-2005 & $\underset{\left(0.245,0.391\right)}{0.316}$ & $\underset{\left(0.389,0.419\right)}{0.404}$ & $\underset{\left(0.031,0.049\right)}{0.041}$ & $\underset{\left(0.049,0.053\right)}{0.051}$\tabularnewline
2005-2006 & $\underset{\left(0.216,0.341\right)}{0.280}$ & $\underset{\left(0.372,0.402\right)}{0.387}$ & $\underset{\left(0.030,0.051\right)}{0.040}$ & $\underset{\left(0.052,0.055\right)}{0.053}$\tabularnewline
2006-2007 & $\underset{\left(0.842,0.930\right)}{0.886}$ & $\underset{\left(0.811,0.845\right)}{0.828}$ & $\underset{\left(0.130,0.163\right)}{0.148}$ & $\underset{\left(0.138,0.142\right)}{0.140}$\tabularnewline
2007-2008 & $\underset{\left(0.866,0.936\right)}{0.900}$ & $\underset{\left(0.854,0.880\right)}{0.868}$ & $\underset{\left(0.094,0.126\right)}{0.111}$ & $\underset{\left(0.103,0.107\right)}{0.106}$\tabularnewline
2008-2009 & $\underset{\left(0.220,0.346\right)}{0.283}$ & $\underset{\left(0.340,0.372\right)}{0.356}$ & $\underset{\left(0.043,0.065\right)}{0.054}$ & $\underset{\left(0.061,0.065\right)}{0.063}$\tabularnewline
2009-2010 & $\underset{\left(0.210,0.322\right)}{0.267}$ & $\underset{\left(0.329,0.360\right)}{0.344}$ & $\underset{\left(0.040,0.062\right)}{0.051}$ &  $\underset{\left(0.060,0.065\right)}{0.062}$\tabularnewline
2010-2011 & $\underset{\left(0.262,0.380\right)}{0.321}$ & $\underset{\left(0.361,0.390\right)}{0.375}$ & $\underset{\left(0.032,0.053\right)}{0.043}$ & $\underset{\left(0.050,0.054\right)}{0.052}$\tabularnewline
2011-2012 & $\underset{\left(0.230,0.356\right)}{0.294}$ & $\underset{\left(0.367,0.3967\right)}{0.381}$ & $\underset{\left(0.028,0.048\right)}{0.038}$ & $\underset{\left(0.049,0.054\right)}{0.051}$\tabularnewline
2012-2013 & $\underset{\left(0.187,0.307\right)}{0.247}$ & $\underset{\left(0.338,0.370\right)}{0.354}$ & $\underset{\left(0.034,0.055\right)}{0.045}$ & $\underset{\left(0.058,0.063\right)}{0.061}$\tabularnewline
\hline
\end{tabular}
\end{table}

\begin{table}[H]
\caption{Estimates of transition probabilities estimates of the mixture of the Gaussian, Gumbel and Clayton copulas taking into account the point masses at zero incomes and 95\% credible intervals (in brackets) \label{tab:Transition-Probability aug-1-1}}
\centering{}%
\begin{tabular}{ccccc}
\hline
Transition & \multicolumn{2}{c}{Stay at 0} & \multicolumn{2}{c}{Stay at positive wages}\tabularnewline
\hline
\hline
 & Non-Parametric & Copula & Non-Parametric & Copula\tabularnewline
\hline
2001-2002 & $\underset{\left(0.469,0.612\right)}{0.535}$ & $\underset{\left(0.493,0.523\right)}{0.508}$ & $\underset{\left(0.950,0.970\right)}{0.960}$ & $\underset{\left(0.955,0.958\right)}{0.956}$\tabularnewline
2002-2003 & $\underset{\left(0.566,0.710\right)}{0.639}$ & $\underset{\left(0.577,0.608\right)}{0.593}$ & $\underset{\left(0.945,0.965\right)}{0.955}$ & $\underset{\left(0.948,0.952\right)}{0.950}$\tabularnewline
2003-2004 & $\underset{\left(0.623,0.766\right)}{0.687}$ & $\underset{\left(0.587,0.617\right)}{0.602}$ & $\underset{\left(0.948,0.968\right)}{0.957}$ & $\underset{\left(0.946,0.949\right)}{0.948}$\tabularnewline
2004-2005 & $\underset{\left(0.610,0.755\right)}{0.684}$ & $\underset{\left(0.581,0.612\right)}{0.597}$ & $\underset{\left(0.951,0.970\right)}{0.960}$ & $\underset{\left(0.947,0.951\right)}{0.949}$\tabularnewline
2005-2006 & $\underset{\left(0.659,0.784\right)}{0.720}$ & $\underset{\left(0.598,0.628\right)}{0.613}$ & $\underset{\left(0.950,0.970\right)}{0.960}$ & $\underset{\left(0.945,0.949\right)}{0.947}$\tabularnewline
2006-2007 & $\underset{\left(0.070,0.158\right)}{0.114}$ & $\underset{\left(0.155,0.189\right)}{0.172}$ & $\underset{\left(0.837,0.871\right)}{0.852}$ & $\underset{\left(0.857,0.862\right)}{0.860}$\tabularnewline
2007-2008 & $\underset{\left(0.064,0.134\right)}{0.010}$ & $\underset{\left(0.120,0.146\right)}{0.133}$ & $\underset{\left(0.874,0.906\right)}{0.889}$ & $\underset{\left(0.892,0.897\right)}{0.894}$\tabularnewline
2008-2009 & $\underset{\left(0.654,0.780\right)}{0.717}$ & $\underset{\left(0.628,0.660\right)}{0.644}$ & $\underset{\left(0.935,0.957\right)}{0.946}$ & $\underset{\left(0.935,0.939\right)}{0.937}$\tabularnewline
2009-2010 & $\underset{\left(0.678,0.791\right)}{0.733}$ & $\underset{\left(0.640,0.671\right)}{0.656}$ & $\underset{\left(0.938,0.960\right)}{0.949}$ & $\underset{\left(0.935,0.940\right)}{0.938}$\tabularnewline
2010-2011 & $\underset{\left(0.620,0.738\right)}{0.679}$ & $\underset{\left(0.610,0.639\right)}{0.625}$ &  $\underset{\left(0.947,0.968\right)}{0.957}$ & $\underset{\left(0.946,0.950\right)}{0.948}$\tabularnewline
2011-2012 & $\underset{\left(0.644,0.770\right)}{0.706}$ & $\underset{\left(0.604,0.633\right)}{0.619}$ & $\underset{\left(0.952,0.972\right)}{0.962}$ & $\underset{\left(0.946,0.951\right)}{0.949}$\tabularnewline
2012-2013 &  $\underset{\left(0.693,0.813\right)}{0.753}$ & $\underset{\left(0.631,0.662\right)}{0.646}$ & $\underset{\left(0.945,0.966\right)}{0.955}$ & $\underset{\left(0.937,0.942\right)}{0.939}$\tabularnewline
\hline
\end{tabular}
\end{table}

\begin{figure}[H]
\caption{Histogram of real individual wages (\$) for Australia in 2001, 2002,
2003, and 2004 respectively from left to right\label{fig:Histogram-of-real}}

\begin{centering}
\includegraphics[width=15cm,height=8cm]{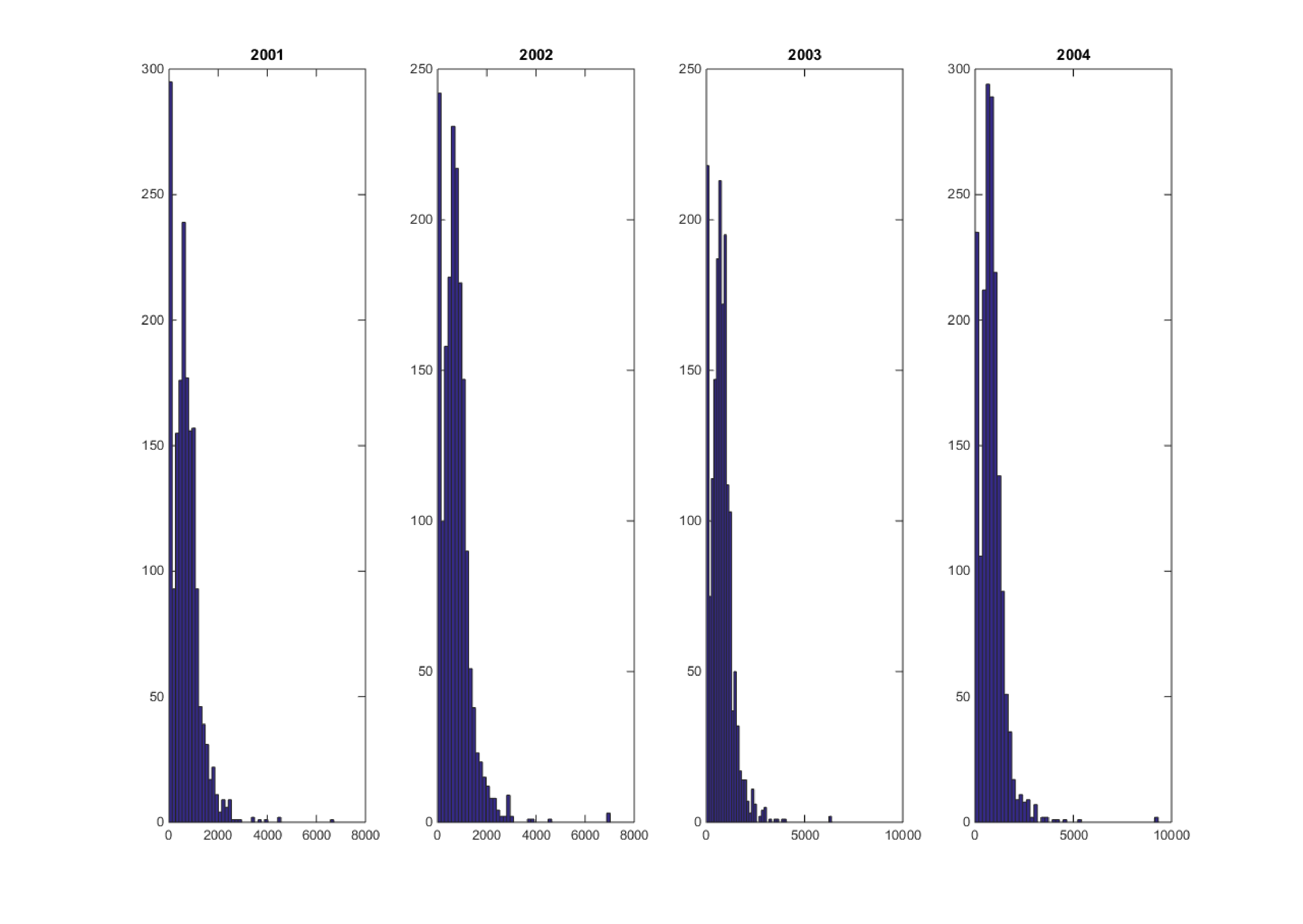}
\par\end{centering}

\end{figure}

\begin{figure}[H]
\caption{Histogram of real individual wages (\$) for Australia in 2005, 2006,
2007, and 2008 respectively from left to right \label{fig:Histogram-of-real-1}}

\centering{}\includegraphics[width=15cm,height=8cm]{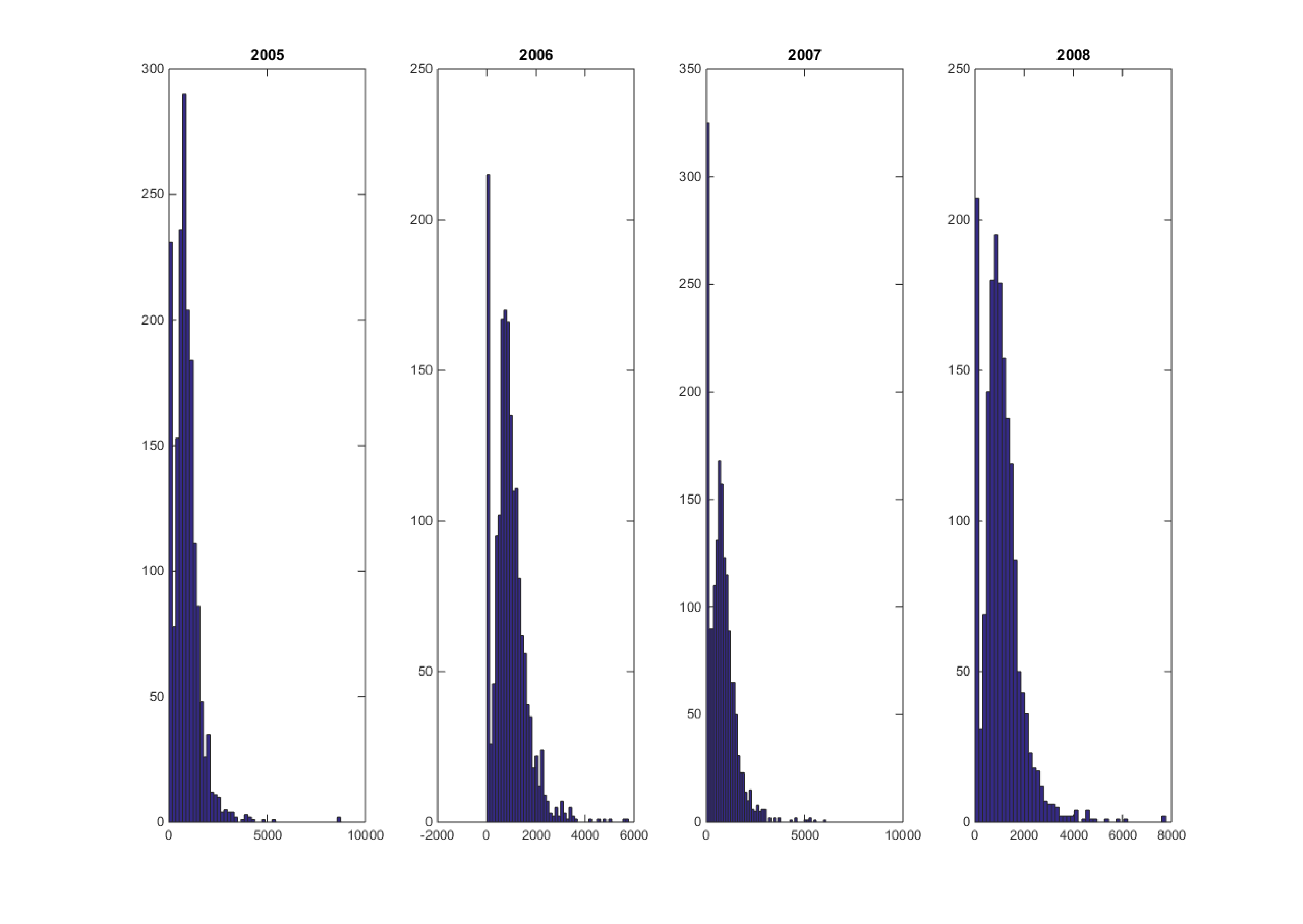}
\end{figure}

\begin{figure}[H]
\caption{Histogram of real individual disposable wages (\$) for Australia in
2009, 2010, 2011, 2012, and 2013 respectively from left to right\label{fig:Histogram-of-real-2}}

\centering{}\includegraphics[width=15cm,height=8cm]{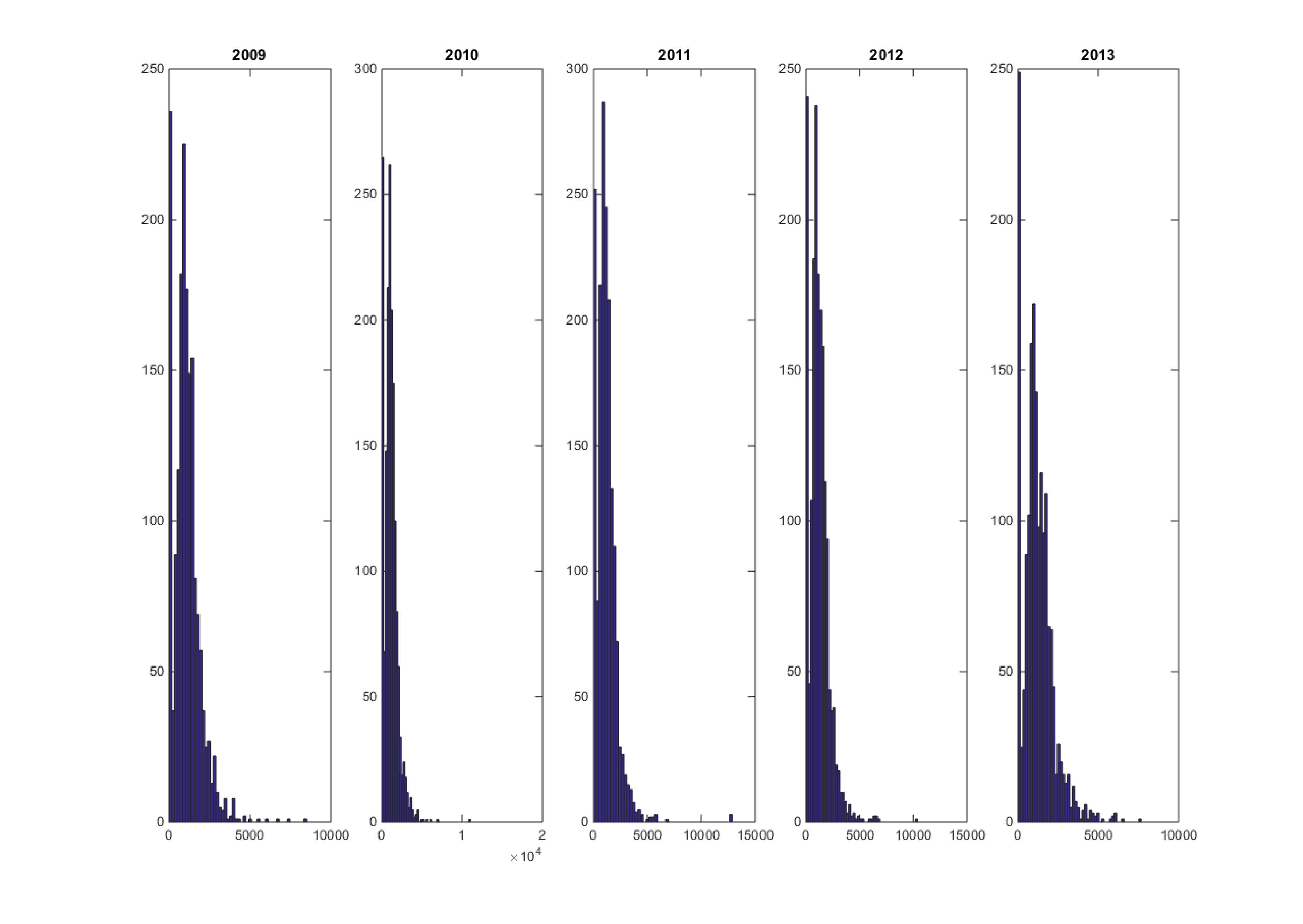}
\end{figure}

\end{document}